\documentclass[pra,twocolumn,aps,superscriptaddress,footinbib]{revtex4-1}
\usepackage{xspace,amsmath,amsfonts,amssymb,amsbsy}
\usepackage{ntheorem}
\usepackage{graphicx,epstopdf}
\usepackage{bbm}
\usepackage{footnote}

\usepackage[breaklinks=true]{hyperref}
\hypersetup{
colorlinks   = true, %Colours links instead of ugly boxes
urlcolor     = blue, %Colour for external hyperlinks
linkcolor    = blue, %Colour of internal links
citecolor    = red %Colour of citations
}
\usepackage{soul}
\usepackage{color}
\usepackage[usenames,dvipsnames]{xcolor}
\usepackage{comment}
 
\usepackage{cleveref} % Must be loaded after hyper ref
\crefformat{equation}{Eq.~(#2#1#3)} % These change 'equation' to Eq., more PRA-style
\crefmultiformat{equation}{Eqs.~(#2#1#3)}{ and~(#2#1#3)}{, (#2#1#3)}{ and~(#2#1#3)}
\Crefformat{equation}{Equation~(#2#1#3)}
\crefformat{section}{Sec.~#2#1#3}
\Crefformat{section}{Section~#2#1#3}
\crefformat{figure}{Fig.~#2#1#3}
\Crefformat{figure}{Figure~#2#1#3}

\newcommand{\sq}[1]{\left[ {#1} \right]}
\newcommand{\tr}[1]{{\textrm {Tr}}\!\sq{#1}}
\newcommand{\smallfrac}[2]{\mbox{$\frac{#1}{#2}$}}
\newcommand{\half}{\smallfrac{1}{2}}
\newcommand{\bra}[1]{\langle{#1}|}
\newcommand{\ket}[1]{|{#1}\rangle}
\newcommand{\ip}[2]{\langle{#1}|{#2}\rangle}
\newcommand{\op}[2]{\ket{#1}\bra{#2}}
\newcommand{\expt}[1]{\langle{#1}\rangle}
\newcommand{\dg}{^\dagger}

\newcommand{\Cexpt}[1]{ {\mathbbm{E}}\! \left [{#1}\right ] }
\newcommand{\nn}{\nonumber}

\newcommand{\sch}{Schr\"odinger}

\newcommand{\Tr}{\mbox{Tr}}

\newcommand{\condOp}{ C_{ \mathbf{R}} }
\newcommand{\mRec}{ \mathbf{R} }
\newcommand{\dt}{_{t}}
\newcommand{\pr}[1]{\Pr( {#1}) }

\newcommand{\MEop}{ \mathcal{L}_{m,n} [ \mathbf{G} ,\xit ]  } 
\newcommand{\Hop}[1]{ \mathcal{H}_{m,n} [ \mathbf{G},\xit, #1 ] } 
\newcommand{\HomCurr}{ K_\phi }
\newcommand{\HomRV}{ d J }

\newcommand{\pie}{_{[0,t)}}         % past
\newcommand{\nie}{_{[t,t+dt)}}      % now
\newcommand{\fie}{_{[t+dt,\infty)}} % future
\newcommand{\pis}{_{t)}}    % past
\newcommand{\nis}{_{t}}     % now
\newcommand{\fis}{_{[t+dt}} % future

\newcommand{\zerot}{0_t}
\newcommand{\onet}{1_t}

\newcommand{\pmt}{\pm_t}

\newcommand{\xit}{\xi(t)}
\newcommand{\xis}{\xi(s)}
\newcommand{\xits}{\xi^*(t)}

\newcommand{\mmvac}{\ket{ \mathbf{0} }}

\newcommand{\alphatot}{\alpha_\xi} % total wavepacket i.e. exp[B(\alapha)-B\dg(alpha]|vac>

\newcommand{\cutout}[1]{\overline{#1}_\xi} % Fock state with current time "cut out"
\newcommand{\het}{ \pm, \widetilde{\pm}_t} % heterodyne eigenstate
\newcommand{\Id}{I}
\newcommand{\dLami}{d\Lambda_t}
\newcommand{\dBi}{dB_t }
\newcommand{\dBdi}{dB_t^{\dag} }
\newcommand{\dLamo}{d\Lambda_t^{\rm out}}
\newcommand{\dBo}{dB_t^{\rm out} }

\newcommand{\Uinf}{U_{t}}
\newcommand{\Hs}{H_{\rm sys}}
\newcommand{\piC}[2]{ \pi_{#1} ( #2 ) }

\newcommand{\Pvac}{\Pi_{\emptyset}} % infinitesimal projector onto vacuum
\newcommand{\Pjump}{\Pi_{J}}   	   % infinitesimal projector onto a single photon
\newcommand{\Phom}{\Pi_{\pm}} 	   % infinitesimal projector onto homodyne eigenstates

\makeatletter
\def\@bibdataout@aps{%
 \immediate\write\@bibdataout{%
  @CONTROL{%
   apsrev41Control,author="08",editor="1",pages="0",title="0",year="1",eprint="1"%
  }%
 }%
 \if@filesw
  \immediate\write\@auxout{\string\citation{apsrev41Control}}%
 \fi
}%
\makeatother % Phew.

\begin{document}

\title{Quantum trajectories for propagating Fock states}

\author{Ben Q. Baragiola}
\email{ben.baragiola@gmail.com}
\affiliation{Centre for Engineered Quantum Systems, Macquarie University, Sydney, NSW 2109, Australia}
\affiliation{Center for Quantum Information and Control, University of New Mexico, Albuquerque, NM 87131, USA}
\author{Joshua Combes}
\email{joshua.combes@gmail.com}
\affiliation{Center for Quantum Information and Control, University of New Mexico, Albuquerque, NM 87131, USA}
\affiliation{Institute for Quantum Computing and Department of Applied Mathematics, University of Waterloo, Waterloo, ON N2L 3G1, Canada}
\affiliation{Perimeter Institute for Theoretical Physics, Waterloo, ON N2L 2Y5, Canada}
\affiliation{Centre for Engineered Quantum Systems, School of Mathematics and Physics, University of Queensland, Brisbane, QLD 4072, Australia}

\date{27 July 2017}

\begin{abstract}
We derive quantum trajectories (also known as stochastic master equations) that describe an arbitrary quantum system probed by a propagating wave packet of light prepared in a continuous-mode Fock state. 
We consider three detection schemes of the output light: photon counting, homodyne detection, and heterodyne detection. We generalize to input field states in superpositions and mixtures of Fock states and illustrate our formalism with several examples.
\end{abstract}

% Quantum information, 03.67.-a
% Light interaction with matter, 42.50.Ct
% Nonlinear optics, 42.65.-k
% Quantum noise, 42.50.Lc
% Fluctuation phenomena quantum optics, 42.50.Lc
% Quantum optics, 42.50.-p
% Antibunched photon states, 42.50.Dv
%    03.65.Yz - Decoherence; open systems; quantum statistical methods

\maketitle

%=====================================================
\section{Introduction}
%=====================================================

Propagating Fock states, wave packets with a definite number of photons, are well suited for the role of relaying information between nodes of a quantum computing device \cite{ReisRemp15}.  In the optical and microwave domains, single-photon fields are routinely produced and manipulated, with ongoing progress toward higher photon numbers \cite{HofhWeigAnsm08,PeauSayrZhou13,CoopWrigSoll13}.  
Taking advantage of Fock states for quantum technology necessitates a theoretical understanding of their interaction with fundamental quantum components, \emph{e.g.} an atom coupled to a waveguide or a transmon coupled to a transmission line in superconducting circuit QED. 

Two useful tools for understanding light-matter interactions are master equations and stochastic master equations. Each is an equation of motion for the reduced state of a quantum system that is coupled to an electromagnetic field. After interacting with the system the field propagates away, carrying with it information. The master equation (ME) is a differential equation for the unconditional reduced system state and ignores any information in the field. The information is not gone; however, and may be retrieved by performing measurements of the output field. 
A time-continuous measurement of the output fields can be used to find the conditional state of the system, known as a \emph{quantum trajectory}~\cite{Carm93a,Carm08}. The equation that takes the sequence of measurement results and determines the system state's conditional evolution is called a stochastic master equation (SME).
 Different types of field measurement---for instance, direct photon counting or quadrature measurements---yield different SMEs. 

%=====================================================
%% FIGURE 1: Depiction %%%
\begin{figure}[b]
\begin{center}
\includegraphics[width=0.9\hsize]{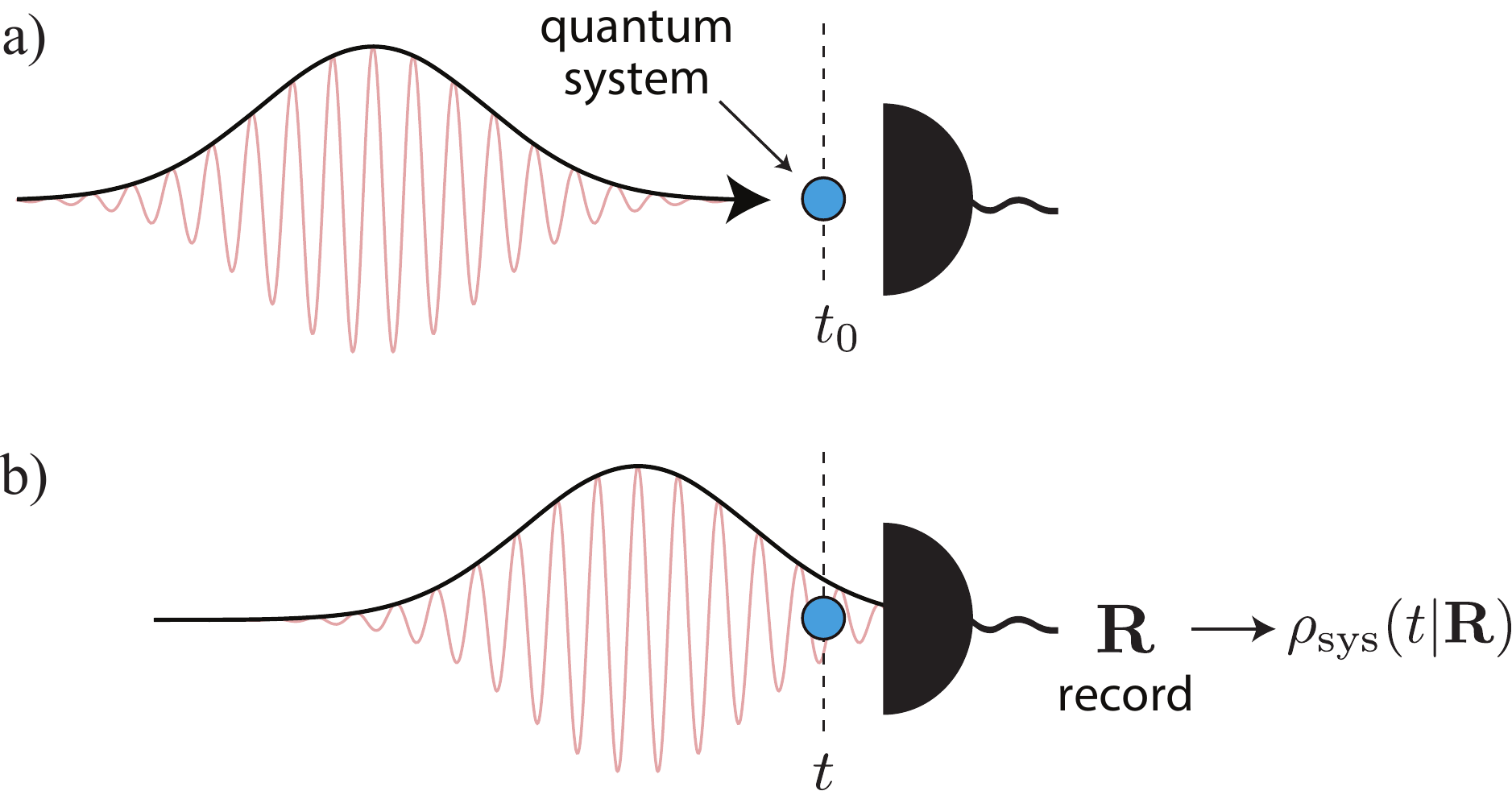}
\caption{
Depiction of a propagating wave packet interacting with a quantum system followed by measurement of the scattered field. The temporal wave packet is given by a slowly-varying envelope $\xit$ modulating fast oscillations at the carrier frequency.  We consider a wave packet prepared in a nonclassical state of definite photon number---a Fock state $\ket{N_\xi}$. a) At time $t_0$, prior to the interaction, the system and field are unentangled. b) At a later time $t>t_0$, a portion of the wave packet has interacted with the system, and the scattered field has been detected. A quantum trajectory describes the conditional reduced state of the system, $\rho_{\rm sys}(t|\mRec)$, given the measurement record $\mRec$. } \label{fig1}
\end{center}
\end{figure}
%=====================================================

The form of an SME depends on the input state of the field. SMEs for Gaussian input fields (vacuum, coherent, thermal, and squeezed) were derived a long time ago \cite{Wise94} and are widely used \cite{WiseMilb10, Murch:2013aa, Campagne-Ibarcq:2016aa, Chantasri:2016aa, Koro16}. Due to the inherently nonclassical nature of Fock states, their SMEs have proven more elusive.    
Propagating Fock states possess temporal correlations which are mapped onto entanglement between system and field as they interact. Consider a single-photon field interacting with a quantum system, see \cref{fig1}.  Classically, two paths have been taken by time $t$: (i) the photon has interacted with the system at some previous time $t' \leq t$, or (ii) the photon has not yet arrived at the system and can be found with certainty in the remaining, future input field $t' > t$.  
The origin of system-field entanglement is the superposition of these options. This is a departure from the typically considered situation for open quantum systems where the input field at time $t$ is uncorrelated both with the system and with the field at all other times.   

In this article we derive stochastic master equations for an arbitrary quantum system probed with a propagating Fock state. 
The temporal correlations in the input field and the entanglement between the system and field are accounted for with a set of coupled equations. 
We consider three different detection schemes of the output light: photon counting, homodyne detection of an arbitrary quadrature, and heterodyne detection. We extend our derivation to include input fields in superpositions or mixtures of Fock states and to inefficient measurements of the field.

Previously some aspects of our theory have been developed. The unconditional MEs for one- and two-photon input fields were introduced by \citet{GherElliPell98} and later extended to many photons in the Fock-state MEs of \citet{BaraCookBrai12}.
A step towards Fock-state SMEs was given by Gheri \emph{et al.} \cite{GherElliPell98}, where they suggested enlarging the Hilbert space to include the source of a single photon using a cascaded approach~\cite{Carm93,Gard93}.
The SMEs for input fields containing any superposition or mixture of vacuum and a single photon were given by \citet{GougJameNurd11,GougJameNurd12}.  Recently, various mathematical techniques have extended these results to SMEs for multiphoton input fields \cite{SongZhanXi16} and for a class of continuous matrix-product states that includes time-ordered, multiphoton states \cite{GougJameNurd14}. 
These derivations proceed in the Heisenberg picture and rely on mathematical techniques unfamiliar to the physics community. 
In contrast we provide a \sch{}-picture derivation, based on the temporal structure of propagating Fock states, that generalizes the Wiseman-Milburn techniques~\cite{Wise94,WiseMilb10}. 
Our approach gives insight into the physical significance of the equations, provides solid footing for useful generalizations, and enables SMEs for any state of the field in a given temporal mode.

 This article is organized as follows.  In \cref{Sec::PhysicalModel} we present the physical model for the interaction of a quantum system with a continuous-mode field using input-output formalism.  The properties of Fock states are summarized with an emphasis on a \emph{temporal decomposition} of the field, and a formal description of continuous field measurements and quantum trajectories is given. In \cref{Sec::SMEFock} we present the main results of this article: the stochastic master equations for quantum systems probe by continuous-mode Fock states when the output fields are subject to three types of measurement: photon counting, homodyne detection, and heterodyne detection. In \cref{Sec::Generalizations} we generalize the SMEs to superpositions and mixtures of Fock states and show how to model imperfect detection and add additional decoherence.  As a pedagogical example, we analyze conditional dynamics of a two-level atom in \cref{Sec::TLSexample}. Finally, we summarize our results and provide outlook in \cref{Sec::Conclusion}.

%=====================================================
\section{Physical model and formalism}  \label{Sec::PhysicalModel}
%=====================================================

	%=====================================================
	\subsection{System-field interaction: input-output theory}  \label{Sec::InputOutput}
	%=====================================================
	
A quantum system interacting with a traveling field naturally calls for a formulation in the time domain due to the time-local nature of the interaction and measurements.  
Input-output theory and the underlying continuous-mode quantization of the field provide such a description \cite{GardZoll04,WiseMilb10}.  
Input-output theory is often formulated for an effectively one-dimensional electromagnetic field, such as arises for one-sided cavities \cite{GardZoll04}, photonic waveguides \cite{Le-KRaus14,CaneManzShi15}, transmission lines in circuit QED \cite{LaluSandLoo13}, and paraxial free-space fields \cite{Cook13}, although this is not a necessary restriction \cite{DumParkZoll92}.  
Within this formalism, enforcement of the weak-coupling limit, the Markov approximation, and the rotating wave approximation yields a quantum stochastic differential equation (QSDE) for the time evolution operator that governs unitary system-field dynamics~\cite{vanHStocMabu05,Cook13}. 

Consider a one-dimensional, continuous-mode field described by bosonic operators satisfying $[b(\omega), b\dg(\omega')] = \delta(\omega - \omega')$.  
For quasi-monochromatic excitations, \emph{i.e.} when the spectral bandwidth is much smaller than the carrier frequency $\Delta \omega \ll \omega_c$, we define slowly varying, Fourier-transformed field operators \cite{MilbBasi15},
	\begin{align} \label{Eq::FieldOps}
		b(t) \equiv \frac{1}{\sqrt{2\pi}} \int_{-\infty}^{\infty} d\omega \, b(\omega) e^{-i (\omega - \omega_c) t}.
	\end{align}
They satisfy the commutation relations $[b(t), b\dg(t') ]=\delta(t-t')$ and are often referred to as white-noise operators, akin to classical white noise, which is $\delta$-correlated in time. 
Here, $t$ is a label for the mode of the field that interacts with the system at time $t$.
Due to the singular nature of $b(t)$ and $b\dg(t)$, it is often preferable to work with the \emph{quantum noise increments} over an infinitesimal time interval $[t, t+dt)$: 
	\begin{subequations} \label{Eq::NoiseIncrements}
	\begin{align}
		&\dBi = \int_t^{t+dt}\!\! ds \, b(s),  \quad \dBdi = \int_t^{t+dt} \!\!ds \, b^\dagger(s),\\
		&\hspace{2cm}	\dLami = \int_t^{t+dt}\!\! ds \, b\dg(s) b(s).
	\end{align}
	\end{subequations}
In the interval $[t, t+dt)$ the operators $dB_t$ and $dB_t\dg$ are infinitesimal annihilation and creation operators, and $d\Lambda_t$ is the infinitesimal number operator. The increments act nontrivially only on $[t,t+dt)$, \emph{e.g.}
\begin{align} \label{Eq::dBfullincrement}
	\dBi = \Id\pie \otimes dB\nie \otimes \Id\fie,
\end{align}
where the interval $[0,t)$ represents the past Hilbert space of the field (the output field), the infinitesimal interval $[t,t+dt)$ the field presently interacting with the system, and the interval $[t+dt, \infty)$ the future field.
  The rules for taking products of the quantum noise increments are given by the It\={o} table \cite{GardZoll04},
	\begin{align} \label{Eq::ItoTable}
		\begin{array}{l}
		\dBi \dBdi = dt,\,\,\, \dBi \dLami = \dBi, \\
		\dLami \dLami = \dLami,\,\,\, \dLami \dBdi= \dBdi,
		\end{array}
	\end{align}
with all other products vanishing to order $dt$. 

Over the infinitesimal time interval $[t, t+dt)$, the time evolution operator \cite{GardZoll04,HudsPart84},
	\begin{align} \label{Eq::InfinitesimalUnitary}
		\Uinf =& \Id_{\rm sys} \otimes \Id\nis - dt \big( i \Hs +\half L^\dagger L \big) \otimes \Id\nis \\
		&- L\dg S \otimes \dBi + L \otimes \dBdi +(S - \Id_{\rm sys}) \otimes \Lambda_t, \nn
	\end{align} 
describes the unitary coupling of the field operators, \cref{Eq::NoiseIncrements}, to system operators $\Hs$, $L$, and $S$, which are determined from the Hamiltonian governing the underlying physical system \cite{Goug06,CombKercSaro16}.  From this interaction, the output relations for the quantum noise increments are \cite{GardColl85,Barc90},
	\begin{subequations}\label{Eq::IO}
	\begin{align}
		\dBo      =& dtL\otimes \Id\nis + S\otimes\dBi, \\
		\dLamo =& dtL\dg L \otimes\Id\nis + L\dg S \otimes\dBi \nn \\
			      & +S\dg L \otimes\dBdi + \Id_{\rm sys} \otimes\dLami.  \label{Eq::dLamOut}
	\end{align}
	\end{subequations}
These are the standard input-output relations \cite{GardZoll04,BoutHandJame07, WiseMilb10} for system Hamiltonian $H_{\rm sys}$ and ``jump operator'' $L$ with the additional inclusion of the scattering operator $S$. Each relation is the coherent sum of two contributions: the free field and system scattering.

	%=====================================================
	\subsection{Continuous-mode Fock states} \label{Sec::FieldStates}
	%=====================================================

Quasi-monochromatic, continuous-mode fields have a convenient description in terms of the input operators of \cref{Eq::FieldOps} \cite{Loud00}.
The utility of this description becomes evident when one considers that interactions, \cref{Eq::InfinitesimalUnitary}, and measurements are local in time.
The continuous-mode field is represented in a continuous temporal tensor-product space, $\mathcal{H}_{\rm field} = \bigotimes_i \mathcal{H}_{t_i}$, where $\mathcal{H}_{t_i}$ is the Hilbert space associated with the field at time $t_i$ \cite{Cook13}.
The statistics of field states that factorize with respect to a given time $t$, 
	\begin{align}
		\ket{ \Psi } = \ket{ \Psi_{t)} } \otimes \ket{ \Psi_{[t} },
	\end{align}
are specified independently on each time interval--the \emph{past} $[t_0,t)$ and the \emph{future} $[t, \infty)$. Field states that factorize temporally with respect to \emph{all} times are described by a temporal tensor-product state,
	\begin{align}
		\ket{ \Psi } = \bigotimes_i \ket{ \Psi_{ t_i} },
	\end{align}
whose statistics can be independently specified over any time interval. 
Critical to the derivation of a master equation or SME is whether the input field is described by a temporal tensor-product state.  

The elementary continuous-mode state is vacuum, given by the product state,
	\begin{align} 
		\mmvac \equiv \bigotimes_{i} \ket{0_{t_i}} = \ket{0} \otimes \ket{0} \otimes \dots,
	\end{align}
indicating vacuum at every time. 
From vacuum, continuous-wave coherent states are constructed by displacing each vacuum component in $\mmvac$ to the same amplitude; $\ket{\alpha} = \bigotimes_{i} \ket{\alpha_{t_i}}$ \cite{Loud00}.  
 
Field states that do not factorize in time are inherently nonstationary and can possess temporal correlations. We now construct normalized, nonstationary continuous-mode fields in a single temporal mode, with particular emphasis on propagating Fock states. Consider a quasi-monochromatic field in temporal mode $\tilde{\xi}(t) = \xit e^{i \omega_c t}$. Rapid oscillations at the carrier frequency $\omega_c$ are modulated by a slowly varying wave packet $\xit$ that is square-normalized,
	\begin{align} \label{Eq::WPnormalization}
		\int ds |\xis|^2 = 1.
	\end{align}
Field states are constructed in the temporal mode $\xit$ by the wave packet creation operator \cite{Loud00}, 
	\begin{align} \label{Eq::WPcreation}
		B\dg(\xi) \equiv \int ds \, \xis b\dg(s),
	\end{align}  
which satisfies $[B(\xi),B\dg(\xi)] = 1$. For example, a wave-packet coherent state with time-varying amplitude $\alpha(t) = \alpha_0 \xit$, for complex peak amplitude $\alpha_0$, is generated from vacuum by a continuous-mode displacement operator: $\ket{\alpha_\xi} = D[\alpha(t)] \mmvac = \exp[ \alpha_0^* B\dg(\xi) - \alpha_0 B(\xi)] \mmvac$ \cite{Loud00}. 

A propagating single photon in the wave packet $\xit$ is generated by applying the wave packet creation operator, \cref{Eq::WPcreation}, directly to continuous-mode vacuum \cite{Loud00,GherElliPell98}, $\ket{1_{\xi}} = B\dg (\xi) \mmvac.$ 
This can be interpreted as superposition of photon creation times weighted by $\xit$ \cite{BlowLoudPhoe90,Loud00,ChiaGarr08}. 
A straightforward extension provides the definition of continuous-mode Fock states (referred to hereafter as \emph{Fock states}) with $n$ photons \cite{Loud00,BaraCookBrai12}, 
	\begin{align} \label{Eq::Fock}
		\ket{n_{\xi}}  &=\smallfrac{1}{\sqrt{n!}} \big[ B\dg(\xi) \big]^{n} \mmvac,
	\end{align}
satisfying $\ip{m_\xi}{n_\xi} = \delta_{m,n}$. The action of the quantum noise increments in \cref{Eq::NoiseIncrements} on Fock states is \cite{BaraCookBrai12}
	\begin{align}
		\dBi \ket{n_\xi} & = dt \sqrt{n} \xit \ket{n-1_\xi} \label{Eq::dBaction}\\
		\dLami \ket{n_\xi} & = dB_t\dg \sqrt{n} \xit \ket{n-1_\xi}. \label{Eq::dLaction}
	\end{align}

\subsubsection{Temporal decomposition}

Shortly we consider the interaction of Fock states with an arbitrary quantum system.
These interactions are time-local [\cref{Eq::InfinitesimalUnitary}], so it is convenient to perform a \emph{temporal decomposition} of the field state. This decomposition expresses the input Fock state in a basis where its projection on the time of interaction is made explicit.
This is achieved by decomposing the wave-packet creation operator, $B\dg(\xi)$, into three intervals---the past $[t_0,t)$, the current $[t,t+dt)$, the future $[t+dt, \infty)$. 
The current interval is infinitesimal, such that the probability of detecting two photons simultaneously is vanishingly small \cite{Moll68, WiseMilb10,JackColl00} (equivalent to $\dBdi \dBdi = 0$ in the It\={o} table). 
With respect to this decomposition, the wave-packet creation operator in \cref{Eq::WPcreation} is,
	\begin{align} \label{Eq::WavePacketTempDecomp}
		B\dg(\xi) & = \underbrace{ \int_{t_0}^t ds \, \xis b\dg(s)}_{\displaystyle \equiv B\pis \dg(\xi)} + \xit dB_t\dg + \underbrace{\int_{t+dt}^\infty ds \, \xis b\dg(s)}_{\displaystyle \equiv B_{[t+dt }\dg(\xi)},
	\end{align}
where $B\pis \dg(\xi)$ and $B\fis \dg(\xi)$ create photons in the \emph{past} and \emph{future} with respect to $t$. We have used abbreviated interval notation, $[t_0,t) \rightarrow t)$ and $[t,\infty) \rightarrow [t$. Each portion of the wave packet creation operator in \cref{Eq::WavePacketTempDecomp} acts on a different interval of multimode vacuum,
	\begin{align} \label{Eq::VacuumDecomposition}
		\mmvac & = \ket{0 \pis } \otimes \ket{0\dt} \otimes \ket{0 \fis}, 
	\end{align}
where $\ket{0\dt}$ is the infinitesimal vacuum at the current time interval. Applying \cref{Eq::WavePacketTempDecomp} to this decomposition of the vacuum we obtain a temporal decomposition of the single-photon wave packet:
	\begin{align} 
		\ket{1_\xi}  = &\ket{1 \pis } \otimes \ket{0\dt} \otimes \ket{0 \fis} + \sqrt{dt} \xit \ket{0 \pis } \otimes \ket{1\dt} \otimes \ket{0 \fis}\nn  \\
		&+ \ket{0 \pis } \otimes \ket{0\dt} \otimes \ket{1 \fis},
	\end{align}
where the infinitesimal single-photon state in the current time interval is \cite{WiseMilb10}
		\begin{align} \label{Eq::InfSinglePhoton}
		\ket{1\dt} = \frac{\dBdi}{\sqrt{dt}} \ket{0\dt}.
	\end{align}
Inserting \cref{Eq::WavePacketTempDecomp} into the definition of a Fock state in \cref{Eq::Fock} gives the temporal decomposition,	
	\begin{align} \label{Eq::NPhotonDecomposition}
		\ket{n_\xi} & = \smallfrac{1}{\sqrt{n!}} \big[  B\dg\pis(\xi) + \xit \dBdi  + B\dg\fis(\xi) \big]^n \mmvac. 
	\end{align}
Further details about the temporal decomposition of Fock states can be found in Appendix \ref{Appendix:TempDecomp}.

A basis for field states in the interval $[t,t+dt)$ can be constructed from the infinitesmal states $ \ket{0_t}$ and $\ket{1_t}$ \cite[Sec. V A]{JackColl00}.
A relative-state decomposition of an $n$-photon Fock state in this basis yields, 
	\begin{subequations} \label{Eq::KeyRelation}
	\begin{align}
		\ket{n_\xi} &= \ket{0\dt} \ip{0_t}{n_\xi} + \ket{1_t} \ip{1_t}{n_\xi} \\
			&= \ket{0\dt} \otimes \ket{\cutout{n}} + \sqrt{n \, dt} \xit \ket{1_t} \otimes \ket{\cutout{n-1}}, \label{Eq::KeyRelationb}
	\end{align}
		\end{subequations}
which can also be found by expanding \cref{Eq::NPhotonDecomposition}, as shown in Appendix \ref{Appendix:TempDecomp}. The state $\ket{\cutout{n}}$ is the partial projection of a Fock state onto infinitesimal vacuum,
	\begin{align} \label{Eq::CutoutProject}
		\ket{\cutout{n}} & \equiv \ip{0_t}{n_\xi}. %\\
	\end{align}
By taking the inner product of \cref{Eq::KeyRelationb} with itself we relate the inner product of $\ket{\cutout{n}}$  to the original wave packet
	\begin{subequations}\label{Eq:ben_norm}
	\begin{align}
		\ip{\cutout{n}}{\cutout{n}} = & 1 - n dt |\xit|^2 \ip{ \cutout{n-1}}{\cutout{n-1}} \\
			= & 1 - n dt |\xit|^2,
	\end{align}
\end{subequations}
where, in the second line we used $\ip{ \cutout{n-1}}{\cutout{n-1}} = 1 - (n-1) dt |\xit|^2 \ip{ \cutout{n-2}}{\cutout{n-2}} $, and kept terms to order $dt$.

	%=====================================================
	\subsection{Continuous measurement of the field} \label{Sec::ContMeasurements}
	%=====================================================
	
After the system and field interact and become entangled the output field is measured. 
Continuous-time measurements are described by a time-ordered sequence of infinitesimal measurements collected as the field impinges on the detection apparatus.  
The infinitesimal time interval is short enough that the probability of detecting two photons is negligible; other situations are obtained by integrating over time.  
In an infinitesimal interval, a projective measurement with outcome $R$ is described by the partial projector,
	\begin{align} \label{Eq::MeasProj}
		\Pi_R = \Id\pis \otimes \op{R\dt}{R\dt} \otimes \Id\fis,
	\end{align}
where $\ket{R\dt} = a_R \ket{0\dt} + b_R \ket{1\dt}$ with $|a_R|^2 + |b_R|^2 = 1$.
The partial projector in \cref{Eq::MeasProj} and its complement act nontrivially only on the infinitesimal interval where the measurement is performed, $[t, t+dt)$. Together they constitute a two-outcome POVM that resolves the identity on the infinitesimal Fock space, $\sum_{R} \Pi_R= \Id\pis \otimes \Id_t \otimes \Id\fis = \Id_{\rm field}$. 

A sequence of such infinitesimal measurements over a time interval $[t_0,t)$ comprises a continuous measurement. The collection of measurement results $\mRec$, referred to as the measurement record, is described by a partial projector, 
	\begin{align} \label{Eq::FullFieldProj}
		\Pi_{\mRec} = \op{ \mRec }{ \mRec } \otimes I_{[t},
	\end{align}
where $\ket{ \mRec}$ is a tensor product of infinitesimal projective eigenstates, as in \cref{Eq::MeasProj}, 
	\begin{align} \label{Eq::MeasEigenstate}
		\ket{ \mRec} = \ket{R_{[t-dt,t)}}\otimes\cdots \otimes \ket{R_{[{t_0},{t_0}+dt)}}= \bigotimes_{t_i} \ket{R_{t_i}}.
	\end{align}
The partial projectors resolve the identity over the time interval $[t_0,t)$,
	\begin{align} \label{Eq::RecIdentity}
		\int d \mRec \,\op{ \mRec }{ \mRec }= \Id_{t)} ,
	\end{align}
and, with \cref{Eq::FullFieldProj}, over the full Fock space \cite[Sec. V]{JackColl00}. 
Regardless of the specific record $\mRec$, the measured portion of the field becomes disentangled from the the future field.

	%=====================================================
	\subsection{Quantum trajectories} \label{Sec::QuantTraj}
	%=====================================================

Consider an initially unentangled quantum system and continuous-mode field described by the joint state,
	\begin{align} \label{Eq::InitialJointState}
		\rho_{\rm joint}(t_0) = \rho_0 \otimes \op{\Psi}{\Psi},
	\end{align}
where $\rho_0$ is the initial system state and $\ket{\Psi}$ the initial field state.  
The time evolution operator that entangles the system and field in each infinitesimal time interval $[t,t+dt)$ has the general form given in  \cref{Eq::InfinitesimalUnitary}.  The total time evolution from $t_0$ to $t$ is given by the time-ordered product of infinitesimal unitaries, 
	\begin{align}
		U(t_0,t) & = U(t-dt,t) \cdots  U(t_0,t_0+dt) \nn \\
		& =\overleftarrow {\prod_s} U(s, s+dt) . \label{Eq::UnitaryTempDecomp}
	\end{align}
Through unitary evolution, the entangled joint state of the system and field at time $t$ is given by
	\begin{align} 
		\rho_{\rm joint}(t) = U(t_0,t) \rho_{\rm joint}(t_0) U\dg(t_0,t). 
	\end{align}

When measurements are performed on the output field, the joint state is conditioned on the random measurement outcomes. For a measurement record $\mathbf{R}$ described by the projector in \cref{Eq::FullFieldProj}, the conditional joint state is found with the usual measurement update rule
	\begin{align} \label{Eq::FullJointConditional}
		\rho_{\rm joint}(t|\mRec) &= \frac{ \op{ \mathbf{R} }{ \mathbf{R} }U(t_0,t)  \rho_{\rm joint}(t_0) U\dg(t_0,t) \op{ \mathbf{R} }{ \mathbf{R} } }{ \Pr(\mathbf{R}) },
	\end{align} 
where $\pr{\mathbf{R}}$ is the probability of obtaining $\bf{R}$.   
It will become useful to describe this quantum operation on the joint state with a \emph{conditional evolution operator} $\condOp$ that includes the interaction $U(t_0,t)$ and measurements $\ket{ \mRec}$ on the interval $[t_0,t)$,
	\begin{align} \label{Eq::CondOp}
		\condOp \equiv \bra{ \mRec } U(t_0,t) \otimes I_{[t}.
	\end{align}
Then, \cref{Eq::FullJointConditional} can be written  
	\begin{align} \label{Eq::FullJointConditionalCR}
		\rho_{\rm joint}(t|\mRec) = \frac{ \condOp \rho_{\rm joint}(t_0) \condOp\dg }{ \Pr(\mathbf{R}) } \otimes \op{ \mathbf{R} }{ \mathbf{R} },
	\end{align} 
with probability given by
	\begin{align}
		\Pr(\mathbf{R}) = \Tr \big[ \condOp\dg \condOp \rho_{\rm joint}(t_0) \big].
	\end{align}
We note that in many physical situations the measured portion of the field is destroyed by the detection process. 
Destructive measurements can be obtained by tracing over the past field in \cref{Eq::FullJointConditional}, leaving only a classical measurement record and the future input field.  
 
 Jack and Collett employed a joint-state description similar to \cref{Eq::FullJointConditional} to study non-Markovian light-matter interactions \cite{JackColl00}. 
While it provides the full, conditional joint state, such a description is difficult to evaluate in practice due to the immense Hilbert space of the continuous-mode field. 
Here, we present an alternate approach and focus on the reduced system state,
	\begin{align} \label{Eq:reducedrho}
		\rho_{\rm sys}(t) = \Tr_{\rm field} [ \rho_{\rm joint}(t) ],
	\end{align}
as the field is continuously measured. From \cref{Eq:reducedrho} forward, we use simplified notation that does explicitly refer to the entire measurement record $\mRec$, except when necessary.

Typically a stochastic master equation (SME), or \emph{quantum trajectory}, is a way to write the the evolution of the reduced system state as the solution to a stochastic differential equation \cite{Carm93a}:
	\begin{align} \label{Eq::dRho}
	d\rho_{\rm sys}(t) = \frac{\bar{\rho}_{\rm sys}(t+dt|R_t)}{\pr{R_t}} - \rho_{\rm sys}(t).
	\end{align}
The primary mathematical objects in the derivation of SMEs are the \emph{Kraus operators} $M_{R} = \bra{R_{t} }U(t,t+dt)\ket{ \psi_{\rm field} }$. Given measurement result $R_t$, the Kraus operators provide a method for the explicit calculation of the unnormalized conditional state above,  
	\begin{align} \label{eq:RedStateMap}
		\bar{\rho}_{\rm sys}(t+dt|R_t) = M_{R} \rho_{\rm sys}(t) M_{R} \dg ,
	\end{align}
and the probability of outcome $R_t$,
	\begin{align}
		\pr{R_t} =\mbox{Tr} \big[ M_{R} \dg M_{R}  \rho_{\rm sys}(t) \big],
	\end{align}
which normalizes the state.
The reduced-state dynamics are Markovian when the input field factorizes temporally with respect to the interaction and measurements, \emph{e.g.} vacuum and coherent states. 
In this case, the map for the reduced state from time $t$ to $t+dt$, \cref{eq:RedStateMap}, is entirely specified by the joint state at time $t$, the entangling unitary [\cref{Eq::InfinitesimalUnitary}], and the measurement result in that interval, $R_t$.  

For input fields that do not admit a temporal tensor-product factorization, such as  continuous-mode propagating Fock states (\cref{Sec::FieldStates}), the reduced-system dynamics are manifestly non-Markovian, and the SME cannot be written in the form of \cref{Eq::dRho}.  
The major result of this manuscript is a technique to generalize this SME to the case of propagating Fock-state input fields, which we present in \cref{Sec::SMEFock}.

%=====================================================	
\section{Fock-state stochastic master equations} \label{Sec::SMEFock}
%=====================================================

We are now equipped to tackle the primary focus of this manuscript---conditional dynamics for a quantum system probed by a continuous-mode, propagating Fock state. 
Our approach uses a set of coupled stochastic master equations.
Their derivation follows from an extension of the standard approach to deriving SMEs in the \sch{}-picture, which begins by identifying the Kraus operators for  infinitesimal measurements of the field \cite{WiseMilb10}. 

In Sec \ref{Sec::SMEStructure} we describe the mathematical objects that arise in the derivation.  In the subsections that follow we derive the SMEs for photon counting, homodyne, and heterodyne measurements. In each section, we state the result and then provide the derivation for interested readers. The Heisenberg-picture formulation of the Fock-state SMEs is given in Appendix \ref{Appendix::HeiSMEs}.

	\subsection{Structure of the SMEs} \label{Sec::SMEStructure}

We begin with a quantum system and input field in the joint state
		\begin{align} \label{Eq::InitialState}
			\rho_{\rm joint}(t_0) = \rho_0 \otimes \op{N_\xi}{N_\xi},
		\end{align}
with the system initially in state $\rho_0$ and the field described by a propagating $N$-photon Fock state $\ket{N_\xi}$, given by \cref{Eq::Fock}.  
As the wave packet arrives, the system and field become entangled by the infinitesimal unitary, \cref{Eq::InfinitesimalUnitary}. 
Then, the output portion of the field is measured.  
For arbitrary time $t$ the joint state of the system and field is formally given by \cref{Eq::FullJointConditional}, and the reduced system state by
	\begin{align} \label{Eq::ReducedStateFock}
		\rho_{\rm sys}(t) & = \frac{1}{\pr{\mRec} } \Tr_{\rm field} \big[\condOp \rho_0 \otimes \op{N_\xi}{N_\xi} \condOp\dg \big].
	\end{align}
We show below that the SME for the reduced system state $\rho_{\rm sys}(t)$ couples to a family of density-matrix-like operators,
	\begin{align} \label{Eq::ReducedStateFock}
		\rho_{m,n}(t) & \equiv \frac{1}{\pr{\mRec } } \Tr_{\rm field} \big[\condOp \rho_0 \otimes \op{m_\xi}{n_\xi} \condOp\dg \big] 
	\end{align}
for $0 \leq \{m,n\} \leq N$. These operators represent fixed photon-number subspaces and coherences between them. Clearly, for $N$-photon Fock-state input the reduced system state is given by the top-level equation: $\rho_{\rm sys}(t) = \rho_{N,N}(t)$. 
The operators $\rho_{m,n}(t)$, which have the same Hilbert-space dimension as $\rho_{\rm sys}(t)$, first arose in the single- and two-photon master equations of Gheri \emph{et al.} \cite{GherElliPell98} and reappeared for conditional and unconditional reduced-state dynamics in a variety of settings \cite{GougJameNurd11,GougJameNurd12,BaraCookBrai12,SongZhanXi16}.
Their initial conditions ($\condOp = I$) are
	\begin{align} \label{Eq::Rhomn}
		\rho_{m,n}(t_0) = \Tr_{\rm field} \big[ \rho_0 \otimes \op{m_\xi}{n_\xi} \big] = \delta_{m,n} \rho_0 .
	\end{align}
That is, the diagonal operators $(m=n)$ are initialized to the system state $\rho_0$. The off-diagonal operators $(m \neq n)$ are initially zero, are traceless at all times, and satisfy $\rho_{m,n}(t) = \rho\dg_{n,m}(t)$ \cite{GherElliPell98, BaraCookBrai12}. 

The field trace in the definition of $\rho_{m,n}(t)$ can be explicitly taken to find the formal quantum operation described by \cref{Eq::ReducedStateFock}. At time $t$, the past portion of the field has interacted with the system and subsequently been measured while the future portion of the field has not.  We show in Appendix \ref{Appendix:TempDecomp} the decomposition of an input Fock state into a future and past basis of Fock states with respect to a chosen time $t$. These bases are defined over two disjoint temporal modes which together comprise the initial wave packet mode. For a field that has been detected up to time $t$, the field trace can be formally taken using the (normalized) Fock basis over the future wave packet, $\ket{n_{[t}}$,
	\begin{align} 
		\rho_{m,n}(t) & = \frac{1}{\pr{\mRec } } \sum_{n'=0}^\infty \Omega_\mRec^{m,n'} \rho_0 \big(\Omega_\mRec^{n,n'} \big)^\dagger,
	\end{align}
where the Kraus operators are
	\begin{align} \label{Eq::GenKrausOps}
		\Omega_\mRec^{m,n'} = \bra{n'_{[t}} \condOp \ket{m_\xi} = \big(\bra{n'_{[t}} \otimes \bra{\mRec} \big) U(t_0,t) \ket{m_\xi}.
	\end{align} 
Our derivation here does not require these full Kraus operators, but we introduce the machinery here for completeness, noting that it will be used in future work to illustrate system-field correlations.	

In the Fock-state SME derivations that follow, the unnormalized operators,
	\begin{align} \label{Eq::StateUpdateUnnorm}
		\bar \rho_{m,n}(t+dt|R\dt) = \frac{1}{\pr{\mRec } } \Tr_{\rm field} \big[ \mathcal{M}^m_{R}(t) \rho_0 \mathcal{M}^{n\dagger}_{R}(t) \big],
	\end{align}	
are updated in each infinitesimal interval with a quantum operation described by the Fock-state {\em pseudo} Kraus operators, $\mathcal{M}^n_{R}(t)$, defined for each input photon number $n$ as indicated by the superscript. These are found by amending $\condOp$ to include the additional conditional dynamics on the current time interval, given by $\Uinf$ and $\ket{R_t}$, and then acting on an input Fock state $\ket{n_\xi}$: 
	\begin{align}  \label{Eq::GenKraus}
		\mathcal{M}^n_{R}(t) \equiv & \bra{R_t} \Uinf \condOp \ket{n_\xi}.
	\end{align}	
The pseudo Kraus operators are distinct from the Kraus operators, \cref{Eq::GenKrausOps}, because they still have support on the Hilbert space of the future field. However, we will make judicious use of $\mathcal{M}^n_{R}(t)$ in the following derivations of the Fock-state SMEs, and for simplicity we henceforth refer to them as the Fock-state Kraus operators. %keeping in mind their distinction from the genuine Kraus operators when performing the trace over the full field.

In the derivations below, we show that the $\rho_{m,n}(t)$ couple amongst themselves and form a closed, Markovian set of equations which can be solved to ultimately find $\rho_{\rm sys}(t)$.
The differential equations have the form
	\begin{align} \label{Eq::dRhoGen}
		d\rho_{m,n}(t|R\dt) 
		&= \frac{ \bar \rho_{m,n}(t+dt|R\dt)}{\pr{R_t}} - \rho_{m,n}(t),
	\end{align}
where, importantly, each $d\rho_{m,n}(t)$ is a function only of $\rho_{m,n}(t)$, $\rho_{m-1,n}(t)$, $\rho_{m,n-1}(t)$, and $\rho_{m-1,n-1}(t)$ for $m,n \geq 0$. The lowest level equation ($m=n=0$) for $d\rho_{0,0}(t)$ closes and is only a function of $\rho_{0,0}(t)$ 
\footnote{Moreover, $\rho_{m,n}(t)$ vanish for any index below zero. This way all of our superoperators are well defined.}.
As $\rho_{m,n}(t) = \rho\dg_{n,m}(t)$, there are at most $\half (N+1)(N+2)$ independent equations \cite{BaraCookBrai12}.  
	
The probability of obtaining the infinitesimal measurement outcome $R_t$ is given by the conditional expectation value of the infinitesimal projector, \cref{Eq::MeasProj}, with respect to the joint state. Using the Fock-state Kraus operators, it can be written as 
	\begin{align} \label{Eq::FockMeasProb}
\pr{R_t} & = \frac{1}{\pr{\mRec}} \Tr \big[ \mathcal{M}^{N\dagger}_{R}(t) \mathcal{M}^N_{R}(t) \rho_0 \big],
	\end{align}	
for an initial $N$-photon Fock state. Note that in \cref{Eq::dRhoGen} all the $\rho_{m,n}(t)$ are rescaled by \cref{Eq::FockMeasProb} for a given measurement outcome. The result is that only the top-level matrix $\rho_{N,N}(t)$, which is indeed the physical reduced state $\rho_{\rm sys}(t)$, remains normalized to unit trace, while the traces of other $\rho_{m,n}(t)$ may vary. 

The Fock-state SMEs are found by careful manipulation of \cref{Eq::StateUpdateUnnorm} and of the measurement probability, \cref{Eq::FockMeasProb}, such that both can be written entirely in terms of the operators $\rho_{m,n}(t)$. Then \cref{Eq::dRhoGen} describes a closed set of coupled equations that can be solved, and the physical reduced state, \cref{Eq::ReducedStateFock}, extracted. Below, we go through this derivation in detail for the Fock-state photon-counting SME.

	%=====================================================
	\subsection{Fock-state photon-counting SME} \label{Sec::FockPCSME}
	%=====================================================

We begin with the photon-counting SME for a quantum system probed by an $N$-photon Fock state.  
After interacting with the quantum system, the output fields are sent to a photodetector. Let $N(t)$ denote the number of photons detected up to time $t$. In the interval $[t, t+dt)$ the random variable $dN$ counts the number of photons, with at most one photon detected. Thus, $dN$ has outcomes $0$ (vacuum detection) and $1$ (photon detection). The conditional evolution under continuous photon counting is given by the set of coupled SMEs,
	\begin{widetext}
	\begin{align} \label{Eq::FockPCSME} 
		&d\rho_{m,n}(t) = \nn  \\
		&   dt \Big( -i[H_{\rm sys},\rho_{m,n}] - \half \big\{ L\dg L, \rho_{m,n} \big\}_+ - \sqrt{m} \xit  L\dg S \rho_{m-1,n}  - \sqrt{n} \xits \rho_{m,n-1} S \dg L - \sqrt{ mn }|\xit|^2  \rho_{m-1,n-1}\Big) + \pr{J} \rho_{m,n}   \nn  \\
 &+ dN \left ( \frac{  L \rho_{m,n} L\dg  + \sqrt{m}\xit S \rho_{m-1,n} L\dg   +  \sqrt{n}\xits L \rho_{m,n-1} S\dg + \sqrt{ mn }|\xit|^2S\rho_{m-1,n-1} S\dg }{ \pr{J}/dt  } -\rho_{m,n} \right ) , 
	\end{align}
	\end{widetext}
where the $\rho_{m,n}$ are defined in \cref{Eq::ReducedStateFock} (henceforth we suppress the argument $t$, except where necessary), for $m,n \in \{0,...,N\}$. The reduced system state is given at all times by the top-level equation, $\rho_{\rm sys} = \rho_{N,N}$, whose evolution is tied to that of other operators $\rho_{m,n}$. The initial conditions for $\rho_{m,n}$ are given by \cref{Eq::Rhomn}. Conditional expectation values of system operators are taken with respect to the reduced system state as usual; \emph{e.g.} $\mathbbm{E}[ X(t)|\mRec] = \Tr [ X  \rho_{\rm sys}(t) ]$. 

The probability of detecting a photon, the ``jump" probability, in the time interval $[t,t+dt)$ is
	\begin{align} \label{Eq::PhotonJumpProb}
		\pr{J} =& dt \, \Tr \big[ L\dg L \rho_{N,N}   + \sqrt{N}\xit L\dg S \rho_{N-1,N}   \\
		& +  \sqrt{N}\xits S\dg L \rho_{N,N-1}  + N |\xit|^2 \rho_{N-1,N-1}  \big]. \nn
	\end{align}
This probability is, in fact, the conditional expectation value of the infinitesimal output photon number operator \cref{Eq::dLamOut}, $\mathbbm{E}[d\Lambda^{\rm out}_t|\mRec]$. Here, we use the notation $\pr{J}$ with the implicit understanding that the jump probability is conditional and depends on the prior measurement record $\mRec$. 
The detection probability, \cref{Eq::PhotonJumpProb}, is a result of photons radiating from the system (first term), photons in the free field (last term), and interference between the two (remaining terms). It is important to note that the trace of $\rho_{N-1,N-1}$, or indeed of any $\rho_{m,n}$ other than the reduced system state $\rho_{\rm sys}$, is not constrained to be equal to 1 \footnote{This differs from the Fock-state master equations in \cite{BaraCookBrai12}, where all diagonal matrices satisfy $\tr{\rho_{n,n}(t)} = 1$.}. 

One can also write \cref{Eq::FockPCSME} in an alternate form, %favored in the quantum filtering literature~\cite{BoutHandJame07,GougJameNurd12}, as
\begin{widetext}
\begin{align} \label{Eq::FockPCSMEInnovations} 
d\rho_{m,n}(t)
=  & dt \, \MEop  \\
  &+d\mathcal{J}_C(t) \left ( \frac{  L \rho_{m,n} L\dg  + \sqrt{m}\xit S \rho_{m-1,n} L\dg   +  \sqrt{n}\xits L \rho_{m,n-1} S\dg + \sqrt{ mn }|\xit|^2S\rho_{m-1,n-1} S\dg   }
 {  \pr{J}/dt  } -\rho_{m,n} \right ), \nn
\end{align}
\end{widetext}
where 
\begin{align} \label{Eq::PCInnovations}
	d\mathcal{J}_{C}(t) \equiv dN- \pr{J},
\end{align}
is called the \emph{photon-counting innovations}.
We have grouped the system operators into the operator-triple, $\mathbf{G} = ( S, L, H_{\rm sys} )$, and introduced $ \MEop$, which is shorthand notation for a superoperator that acts on a set of $\rho_{m,n}$ operators as follows:
\begin{align} \label{Eq::GenLindblad} 
	\MEop \equiv
		& -i[H_{\rm sys},\rho_{m,n}] + \mathcal{D}_L [\rho_{m,n}]  \\
		& + \sqrt{m}\xit  [S \rho_{m-1,n}, L\dg ]   \nn\\
		& + \sqrt{n}\xits  [L, \rho_{m,n-1} S\dg] \nn \\
		& +\sqrt{mn }|\xit|^2  \big( S\rho_{m-1,n-1} S\dg - \rho_{m-1,n-1} \big), \nn
\end{align}
where the Lindblad superoperator is 
	\begin{align} \label{Eq::Lindblad}
		\mathcal{D}_L[\rho] \equiv L \rho L\dg - \half ( L\dg L \rho + \rho L\dg L).
	\end{align}
Equation (\ref{Eq::GenLindblad}) is simply the unconditional part of the evolution---that is, the Fock-state master equations derived in Ref. \cite{BaraCookBrai12}.

		%=====================================================
		\subsubsection{Derivation}\label{subsub:deriv1}
		%=====================================================

The joint system-field state is initialized at $t_0$ in the unentangled state given by \cref{Eq::InitialState}. At a later time $t > t_0$, continuous photon counting has generated a measurement record $\mRec$. 
The projectors for photon counting in an infinitesimal time interval are constructed from eigenstates of the infinitesimal photon-number operator $d\Lambda_t$. Using Carmichael's notation \cite{Carm93a,Carm08} we label the infinitesimal outcomes as $R_t \in \{\emptyset, J\}$, describing either vacuum or a single photon, respectively. These correspond to eigenstates $\ket{0\dt}$ and $\ket{1\dt}$ [\cref{Eq::InfSinglePhoton}] with respective eigenvalues $\{0, 1\}$. 
The associated infinitesimal measurement projectors, \cref{Eq::MeasProj}, are
	\begin{subequations}\label{Eq::PCprojectors}
	\begin{align} 
		\Pvac = & \Id\pis\otimes \op{\zerot}{\zerot} \otimes \Id\fis , \\
		\Pjump = & \Id\pis\otimes \op{\onet}{\onet} \otimes \Id\fis.
	\end{align}
	\end{subequations}
The conditional joint state, \cref{Eq::FullJointConditional}, is subject to entangling interaction and measurement in the infinitesimal time interval $[t, t+dt)$. Our task now is to find the reduced physical state and associated unnormalized operators $\bar{\rho}_{m,n}$, given by \cref{Eq::StateUpdateUnnorm}, for each of the two outcomes in that interval. 

We begin with the case of vacuum detection ($R_t = \emptyset$). The vacuum Kraus operators from \cref{Eq::GenKraus} are of the form
\begin{align} 
		\mathcal{M}^n_{\emptyset}(t) = &  \bra{0_t} \Uinf \condOp \ket{n_\xi}.
\end{align}	
We insert the entangling unitary, \cref{Eq::InfinitesimalUnitary}, and use the relative state decompostion of the input field with respect to the current time, \cref{Eq::KeyRelation}, and find
	\begin{align} 
		\mathcal{M}^n_{\emptyset}(t) = & \big[ \Id_{\rm sys} - dt \big( iH_{\rm sys} +\half L^\dagger L \big)  \big] \condOp \ket{\cutout{n}} \label{Eq::FockKrausVac} \\
			 &- dt \sqrt{n} \xit L\dg S \condOp \ket{\cutout{n-1}} . \nn 
	\end{align}	
To arrive at this expression, we first simplified products of $dB_t$ and $d\Lambda_t$ using the It\={o} table in \cref{Eq::ItoTable}. 
The remaining quantum noise increments satisfy $[dB_t, \condOp] = [dB_t\dg, \condOp] = [d\Lambda_t, \condOp] = 0$, since they are defined on non-overlapping time intervals. Thus they can be pulled through $\condOp$ and applied directly to the Fock state $\ket{n_\xi}$ via \cref{Eq::dBaction}. 
Finally, the partially projected Fock-states, $\ket{\cutout{n}}=\ip{0_t}{n_\xi}$, are the remnants of projecting onto vacuum [see \cref{Eq::KeyRelation}].

To find the unnormalized conditional operators for vacuum detection, we insert \cref{Eq::FockKrausVac} into \cref{Eq::StateUpdateUnnorm} 
	\begin{align}
		\bar \rho_{m,n} & (t+dt|  \emptyset)  = \frac{1}{\pr{ \mRec }} \Tr_{\rm field} \Big[ \condOp \left( \rho_0  \otimes \op{\cutout{m}}{\cutout{n}} \right) \condOp\dg \nn  \\
			& -i dt \big[ H_{\rm sys}, \condOp \left( \rho_0  \otimes \op{\cutout{m}}{\cutout{n}} \right) \condOp\dg \big] \nn \\
			& - dt \half \big\{ L\dg L, \condOp \left( \rho_0  \otimes \op{\cutout{m}}{\cutout{n}} \right) \condOp\dg \big\}_+ \nn \\
			& - dt \sqrt{m}  \xit L\dg S \condOp \left( \rho_0  \otimes \op{\cutout{m-1}}{\cutout{n}} \right) \condOp\dg \nn \\
			& - dt \sqrt{n} \xits \condOp \left( \rho_0 \otimes \op{\cutout{m}}{\cutout{n-1}} \right) \condOp\dg S\dg L   \Big]. \label{Eq::RhoStateMap}
	\end{align}
While the probability of the prior record, $\pr{\mRec}$, appears, the operators remain unnormalized because the current vacuum-detection probability has not been included. 

The critical final step to the derivation lies in writing \cref{Eq::RhoStateMap} exclusively in terms of input Fock states $\ket{n_\xi}$; that is, those defined over the entire input wave-packet mode. This is done by rearrangement of the relative-state decomposition for an $n$-photon Fock state, given in \cref{Eq::KeyRelation}. Consider the first term in \cref{Eq::RhoStateMap}. Under a field trace this relation gives, 
	\begin{align}
		\Tr_{\rm field} & \big[ \condOp \rho_0 \otimes \op{\cutout{m}}{\cutout{n}} \condOp\dg \big] \nn \\
		= & \Tr_{\rm field} \big[ \condOp \rho_0 \otimes \op{m_\xi}{n_\xi}  \condOp\dg \big]   \label{eq:blah1} \\
		 & -   dt \sqrt{mn}  |\xit|^2 \Tr_{\rm field} \big[ \condOp \rho_0 \otimes \op{\cutout{m-1}}{\cutout{n-1}} \condOp\dg \big]\nn , \\
		 = & \Tr_{\rm field} \big[ \condOp \rho_0 \otimes \op{m_\xi}{n_\xi}  \condOp\dg \big]  \\
		 & -   dt \sqrt{mn}  |\xit|^2 \Tr_{\rm field} \big[ \condOp \rho_0 \otimes \op{m-1_\xi}{n-1_\xi} \condOp\dg \big].  \nn
	\end{align}
The second term in \cref{eq:blah1} is not in terms of the initial wavepacket, so in the second equality we recursively apply the same procedure and keep terms to order $dt$. Now, we use the definition of $\rho_{m,n}(t)$ in \cref{Eq::ReducedStateFock} to get the key relation, 
	\begin{align}
		\frac{1}{\pr{ \mRec }}  \Tr_{\rm field} & \big[ \condOp \rho_0 \otimes \op{\cutout{m}}{\cutout{n}} \condOp\dg \big] \nn  \\
		 = & \rho_{m,n}(t) 
		 -  dt \sqrt{mn} |\xit|^2  \rho_{m-1,n-1}(t) . \label{Eq::TheTrickRewrite}  
	\end{align}
We repeat this procedure for the remaining terms in \cref{Eq::RhoStateMap}, neglecting terms of order $dt^2$. The conditional map can now be written entirely in terms of the operators $\rho_{m,n}$, 
	\begin{align} \label{Eq::PhysStateUnnorm}
		\bar \rho_{m,n} & (t+dt|  \emptyset) =   \rho_{m,n} - dt \sqrt{mn} |\xit|^2 \rho_{m-1,n-1}\nn\\
		&-i dt [ H, \rho_{m,n} ]  - dt \half \big\{ L\dg L, \rho_{m,n} \big\}_+ \nn\\
		  & -dt \sqrt{m}  \xit L\dg S \rho_{m-1,n} - dt \sqrt{n} \xits \rho_{m,n-1} S\dg L .  
	\end{align}	 
	
The operators $\bar \rho_{m,n}  (t+dt|  \emptyset)$ are normalized by the probability of detecting vacuum in the time interval $[t,t+dt)$, \emph{i.e.} $ \rho_{m,n}  (t+dt|  \emptyset)= \bar \rho_{m,n}  (t+dt|  \emptyset)/\pr{\emptyset}$. To obtain the vacuum-detection probability we substitute the top-level Kraus operators, $\mathcal{M}^N_{\emptyset}(t)$, into \cref{Eq::FockMeasProb}, 
	\begin{align} \label{Eq::ProbVac}
\pr{\emptyset}   = & 1-\pr{J}\nn\\
=& 1 - dt \Tr \big[ L\dg L \rho_{N,N} + \sqrt{N}  \xit L\dg S \rho_{N-1,N} \nn\\
		& + \sqrt{N} \xits \rho_{N,N-1} S\dg L + N |\xit|^2 \rho_{N-1,N-1} \big] . 
	\end{align}
The probability of vacuum detection is now expressed entirely in terms of the operators $\rho_{m,n}$. The prior normalization, $\pr{\mRec}^{-1}$, is  absorbed into each of the $\rho_{m,n}$ through their definition, \cref{Eq::ReducedStateFock}. Using the standard Taylor expansion of the denominator to order $dt$ \cite{JacoStec06} we find
\begin{align} \label{Eq::PhysStateUnnorm}
		 \rho_{m,n} & (t+dt|  \emptyset) =   \rho_{m,n} - dt \sqrt{mn} |\xit|^2 \rho_{m-1,n-1}\nn\\
		&-i dt [ H, \rho_{m,n} ]  - dt \half \big\{ L\dg L, \rho_{m,n} \big\}_+ \nn\\
		  & -dt \sqrt{m}  \xit L\dg S \rho_{m-1,n} - dt \sqrt{n} \xits \rho_{m,n-1} S\dg L \nn\\
		  &+ \pr{J}\rho_{m,n}  .
	\end{align}	

The case of photon detection ($R_t = J$) proceeds similarly.  The Fock-state Kraus operator for photon detection is
\begin{align}
\mathcal{M}^n_{J}(t) 
=& \bra{1_t} \Uinf \condOp \ket{n_\xi}\nn\\
= & \sqrt{dt} \big( L \condOp \ket{\cutout{n}} + \sqrt{n} \xit S \condOp \ket{\cutout{n-1}} \big). \label{Eq::FockKrausJump}
\end{align}
These are then applied using \cref{Eq::StateUpdateUnnorm} to find $\bar{\rho}_{m,n}(t+dt|J)$, just as was done for vacuum detection. Indeed, we follow the same procedure using \cref{Eq::TheTrickRewrite} to express the right-hand side entirely in terms of the operators $\rho_{m,n}$, to arrive at the expression,
	\begin{align} 
		\bar{\rho}_{m,n} (t+dt|J)   = & dt L\rho_{m,n}L\dg + dt \sqrt{m} \xit S \rho_{m-1,n} L\dg \label{Eq::FockJumpMap} \nn \\
			  & + dt \sqrt{n} \xits L \rho_{m,n-1} S\dg  \nn \\
			  &+dt \sqrt{mn} |\xit|^2 S \rho_{m-1,n-1} S\dg . 	
	\end{align}
The operators $\bar{\rho}_{m,n} (t+dt|J)$ are normalized by dividing by the probability of detecting a photon, $\pr{J}$, found from \cref{Eq::FockMeasProb} and given explicitly in \cref{Eq::PhotonJumpProb}.

For each infinitesimal measurement outcome $\{\emptyset,J\}$ we have expressed the conditional operators in terms of the $\rho_{m,n}$.   
Consequently, we may write down a set of coupled differential equations following the usual procedure \cite{GoetGrah94, WiseMilb10}. For each outcome we use \cref{Eq::dRhoGen} to get
	\begin{align} \label{Eq::SMEFockVac}
		d\rho_{m,n}  (t|\emptyset)  &=  dt \Big( -i [ H, \rho_{m,n} ] - \half \big\{ L\dg L, \rho_{m,n} \big\}_+ \nn  \\
			 &- \sqrt{m} \xit L\dg S \rho_{m-1,n} - \sqrt{n} \xits \rho_{m,n-1} S\dg L\nn \\
			&-  \sqrt{mn} |\xit|^2 \rho_{m-1,n-1} \Big) +  \pr{J}  \rho_{m,n},   
	\end{align}
and
	\begin{align}\label{Eq::SMEFockJump}
		d\rho_{m,n} (t|J) &=  dt [\pr{J}]^{-1} \Big(  L \rho_{m,n} L\dg\nn  \\
		&   + \sqrt{m}\xit S \rho_{m-1,n} L\dg + \sqrt{n}\xits L \rho_{m,n-1} S\dg   \nn \\
		&+ \sqrt{mn }|\xit|^2S\rho_{m-1,n-1} S\dg \Big) -\rho_{m,n}. 
	\end{align}
The Fock-state SME for photon counting, stated explicitly in \cref{Eq::FockPCSME}, is found by combining the conditional equations for vacuum and photon detection into a single differential equation by introducing a binary random variable $dN$ that satisfies $dN^2 =dN$ and has outcomes $0$ (vacuum detection) and $1$ (photon detection). The conditional evolution is then concisely expressed as
	\begin{align}\label{Eq::FockPCSME_simple}
		d\rho_{m,n}(t) 	= & dN d\rho_{m,n}(t|J) + (1-dN)d\rho_{m,n}(t|\emptyset).
	\end{align} 
When a photon is counted ($dN=1$) the state is updated with \cref{Eq::SMEFockJump}, otherwise ($dN=0$) and it is updated with \cref{Eq::SMEFockVac}.  Before the current infinitesimal measurement is performed, the conditional expectation value, $\Cexpt{dN|\mRec} = 0\times \pr{\emptyset} + 1\times \pr{J}= \pr{J}$, is simply the probability for photon detection. Since $\Cexpt{dN|\mRec}$ is of order $dt$, terms of order $dN dt$ vanish \cite{GardZoll04}, which gives the Fock-state photon-counting SME in \cref{Eq::FockPCSME}. 
	
An alternate way to write \cref{Eq::FockPCSME} is in terms of the \emph{photon-counting innovations}, $ d\mathcal{J}_{C}(t)  \equiv dN- \mathbbm{E}[dN|\mRec]$, which is \cref{Eq::PCInnovations}.
The innovations is the difference between the actual measurement outcome and the expected result and characterizes how much is learned from the measurement.  The Fock-state photon-counting SME in \cref{Eq::SMEFockJump} can be transformed to innovations form, 
\begin{align} \label{Eq::FockPCSMEInnovations} 
		 d \rho_{m,n} & (t)  =dt \MEop  + d\mathcal{J}_C(t) d\rho_{m,n}  (t|J) , 
	\end{align}
where $ \MEop$ is defined in \cref{Eq::GenLindblad}.

Provided that one knows the initial system state \cite{CookRiofDeut14}, ensemble averaging over the Fock-state photon-counting SME over trajectories yields the unconditional Fock-state master equations in Ref. \cite{BaraCookBrai12}.  In innovations form this is evident upon inspection since $\mathbbm{E} [d\mathcal{J}_C(t)] = 0$ by definition.

	%=====================================================
	\subsection{Fock-state homodyne SME} \label{Sec::FockHomodyne}
	%=====================================================
	
In homodyne detection, the output field is combined on a balanced beamsplitter with a local oscillator of phase $\phi$.
The two fields exiting the beamsplitter are sent to photodetectors, whose subtracted photocurrents give the homodyne signal. The conditional evolution under homodyne detection is given by the set of coupled SMEs,
\begin{align} \label{Eq::FockHomSME}
	d\rho_{m,n}(t)  = dt \MEop+d\mathcal{J}_\phi(t) \Hop{\phi}, 
\end{align}
where $ \MEop$,  \cref{Eq::GenLindblad}, describes the unconditional evolution, and $\Hop{\phi}$ is a generalization of Wiseman's conditioning map \cite{Wise94a},
\begin{align} \label{Eq::Hop}
	& \Hop{\phi}  \equiv  e^{-i \phi}L\rho_{m,n}  + e^{i \phi} \rho_{m,n} L\dg  \\ 
		& + e^{-i \phi} \sqrt{m}\xit  S\rho_{m-1,n} + e^{i \phi} \sqrt{n}\xits \rho_{m,n-1} S\dg - \HomCurr \rho_{m,n}. \nn 
\end{align}
The \emph{expected homodyne current} is given by
	\begin{align} \label{Eq::QuadCurrent}
		\HomCurr = & \Tr \big[ (e^{-i \phi}L  + e^{i \phi} L\dg ) \rho_{N,N} \\
		& + e^{-i \phi} \sqrt{N}\xit  S\rho_{N-1,N} + e^{i \phi} S\dg \sqrt{N}\xits \rho_{N,N-1} \big], \nn
	\end{align}  
which is equivalent to $\Tr [\mathcal{H}_{N,N}[\mathbf{G},\xit]]$.
The homodyne innovations $d\mathcal{J}_\phi(t)$, 
	\begin{align} \label{Eq::HomInnovations}
		d\mathcal{J}_\phi(t)
		& = dJ_\phi  - dt \HomCurr,
	\end{align}
satisfy the properties of a classical Wiener process: zero mean, $\Cexpt{d\mathcal{J}_\phi(t)} = 0$, and variance $\Cexpt{d\mathcal{J}_\phi(t)^2} = dt$.

		%=====================================================
		\subsubsection{Derivation} \label{subsub:deriv2}	
		%=====================================================

Now that we have a thorough derivation of the Fock-state SME for photon counting, we present the homodyne SME for a quantum system probed by an $N$-photon Fock state with fewer details, as much of the derivation proceeds in exactly the same way.	

The projectors for balanced homodyne measurement in an infinitesimal time interval are constructed from eigenstates of an infinitesimal quadrature operator \cite{GoetGrah94,JackColl00},
	\begin{align} \label{Eq::InfQuadrature}
		dQ_\phi = e^{-i\phi}\dBi+e^{i\phi}\dBdi.
	\end{align}  
The measurement outcomes, labeled by $R_t \in \{ \pm \}$, correspond to eigenstates that are equal superpositions, 
	\begin{align} \label{Eq::HDeigenstates}
		\ket{\pmt} = \smallfrac{1}{\sqrt{2}} \big(\ket{\zerot} \pm e^{i\phi}\ket{\onet} \big ),
	\end{align} 
with eigenvalues given by $dQ_\phi \ket{\pmt} = \pm \sqrt{dt} \ket{\pmt}$~ \cite{GoetGrah94,JackColl00}.
The associated infinitesimal measurement projectors are
	\begin{align}  \label{Eq::HDProjectors}
		\Phom = & \Id\pis\otimes \op{\pmt}{\pmt} \otimes \Id\fis. 
	\end{align}
Modeling continuous homodyne measurement as a series of two-outcome measurements is a straightforward consequence of performing infinitesimal measurements in the single-photon sector \cite{GoetGrah94,JackColl00}. 

The homodyne Kraus operators are obtained from \cref{Eq::GenKraus} and \cref{Eq::HDProjectors}. 
They can be conveniently written as superpositions of the photon-counting Kraus operators, \cref{Eq::FockKrausVac} and \cref{Eq::FockKrausJump}:
	\begin{align}
		\mathcal{M}^n_\pm(t) 
			& = \smallfrac{1}{\sqrt{2}} \left[ \mathcal{M}^n_{\emptyset}(t) \pm  e^{-i \phi} \mathcal{M}^n_{J}(t)  \right].		
	\end{align} 
The probabilities for the outcomes follow from the Kraus operators [\cref{Eq::FockMeasProb}], 
	\begin{align} \label{Eq::HomProb}
		\pr{\pm}
			&= \smallfrac{1}{2}  \big( 1   \pm \sqrt{dt} \HomCurr \big), 
	\end{align}
and satisfy $\pr{+} + \pr{-}=1$. We apply the Kraus operators to find the SME for each $\rho_{m,n}$, just as was done for the case of photon counting above. After expanding \cref{Eq::StateUpdateUnnorm}, we use the relation in \cref{Eq::TheTrickRewrite} to rewrite the expression entirely in terms of the $\rho_{m,n}$ matrices. Then, the unnormalized conditional operators are
	\begin{align} 
  \bar \rho_{m,n}   (t+  dt  |\pm)  
		  =& \,  \half \big\{ \rho_{m,n} + dt \MEop \label{Eq::HomMap}  \\
	&	 \pm  \sqrt{dt} \big[ \Hop{\phi} + \HomCurr  \rho_{m,n} \big] \big\}  ,  \nn
	\end{align}
where the homodyne conditioning map $\Hop{\phi}$ and expected homodyne current $\HomCurr$ are given in Eqs. (\ref{Eq::Hop}-\ref{Eq::QuadCurrent}). 

The conditional states, \cref{Eq::HomMap}, and probabilities, \cref{Eq::HomProb}, are combined into differential equations for each $\rho_{m,n}$ using \cref{Eq::dRhoGen} by expanding the denominator to order $dt$ and collecting terms. We introduce a random variable $\HomRV_\phi$ that takes on values $\pm \sqrt{dt} $. From the conditional expectation values $\Cexpt{\HomRV_\phi |\mRec}= \sqrt{dt} \times \pr{+} - \sqrt{dt} \times \pr{-} = dt \HomCurr$  
and $\mathbbm{E}[\HomRV_\phi^2|\mRec ] = dt $, the variance of $\HomRV_\phi$ is $dt$. 
Thus we define the \emph{homodyne innovations} as 
$ d\mathcal{J}_\phi(t)  \equiv \HomRV_\phi  - \mathbbm{E} [\HomRV_\phi|\mRec] $ which is equivalent to \cref{Eq::HomInnovations}.
As stated above, this satisfies the properties of a classical Wiener process: mean zero and variance $dt$. After some algebra, the differential updates corresponding to each measurement are combined into the Fock-state homodyne SME, \cref{Eq::FockHomSME}.

	%=====================================================
	\subsection{Fock-state heterodyne SME}
	%=====================================================
	
Continuous heterodyne detection ~\cite{Milb87,GoetGrah94,Wise95,Wise96,WiseMilb10} simultaneously measures two orthogonal quadratures, either by mixing the output fields with a detuned local oscillator or performing double homodyne detection. For orthogonal quadratures specified by the phases $\phi \in \{0, \pi/2\}$, the Fock-state heterodyne SME is
	\begin{align} \label{Eq::FockHetSME}
	d\rho_{m,n}(t)  = & \,
		dt \MEop \nn \\
&		+ \smallfrac{1}{\sqrt{2}} d\mathcal{J}_0(t) \Hop{0}  \nn\\
&  		+ \smallfrac{1}{\sqrt{2}}d\mathcal{J}_{\pi/2}(t) \Hop{\pi/2}, 
	\end{align}
where $ \MEop$, \cref{Eq::GenLindblad}, describes the unconditional evolution, and $\Hop{\phi}$ is the conditioning map for each quadrature, \cref{Eq::Hop}.  
Each quadrature has a homodyne innovations 
$\mathcal{J}_\phi(t)$, \cref{Eq::HomInnovations}, satisfying the properties of a classical Wiener process.

		%=====================================================
		\subsubsection{Derivation}	
		%=====================================================
		
As measurements of orthogonal quadratures do not commute, heterodyne detection is a more general measurement of the field described by 
a positive-operator valued measure (POVM). 	

In an infinitesimal time interval each quadrature is described by \cref{Eq::InfQuadrature} with phases $\phi \in \{0, \pi/2\}$, respectively. Each quadrature measurement has two outcomes, which are not independent, giving the four joint outcomes $\{+ \widetilde{+}, + \widetilde-, - \widetilde+, - \widetilde- \}$. The Kraus operators of the infinitesimal field measurement are 
	\begin{align}  \label{Eq::HetProjectors}
		\Upsilon_{\pm,\widetilde{\pm}} = & \Id\pis\otimes \frac{\op{\pm, \widetilde{\pm}_t}{\pm, \widetilde{\pm}_t}}{\sqrt{2}}\otimes \Id\fis, 
	\end{align}
which are composed of the four nonorthogonal states,
	\begin{align} \label{Eq::Heteigenstates}
		\ket{\pm, \widetilde{\pm}_t}&= \frac{1}{\sqrt{2}} \bigg( \ket{0\dt} +  \frac{\pm 1 \widetilde\pm i} {\sqrt{2}}\ket{1\dt} \bigg ). 
	\end{align}  
The POVM elements corresponding to the outcomes are $E_{\pm, \widetilde\pm} =\Upsilon_{\pm,\widetilde{\pm}}\dg \Upsilon_{\pm,\widetilde{\pm}}= \half\op{\het}{\het}$ and obey $\sum_{s,r}E_{s,r}=\Id_t$. 

The heterodyne Kraus oprators can be written in terms of the photon-counting Kraus operators, \cref{Eq::FockKrausVac} and \cref{Eq::FockKrausJump},
	\begin{align}
		\mathcal{M}^n_{\pm, \widetilde\pm}(t) 
		=& \frac{1}{2} \Big[ \mathcal{M}_\emptyset^n(t) + \smallfrac{1}{\sqrt{2}}(\pm1 \widetilde{\pm} i ) \mathcal{M}_J^n(t) \Big].
	\end{align}
The outcomes probabilities follow from \cref{Eq::FockMeasProb}, 
	\begin{align}
		\pr{\pm,\widetilde{ \pm} } =& \frac{1}{4} \Big[ 1 + \sqrt{\smallfrac{dt}{2} } \big( \pm K_{0} \widetilde{\pm} K_{\pi/2} \big) \Big]
	\end{align}
where $\HomCurr$ are the quadrature currents, \cref{Eq::QuadCurrent}.

For each measured quadrature we define a random variable, $\HomRV_0 = \pm \sqrt{dt}$ and $\HomRV_{\pi/2} = \widetilde{\pm} \sqrt{dt}$, which together satisfy the property $\HomRV_i \HomRV_j = dt \delta_{i,j}$ so that $[(\HomRV_i + \HomRV_j )/\sqrt{2}]^2=d t$. The statistics are found from the marginal probability distributions, $	\pr{\widetilde{ \pm} } =	\pr{-,\widetilde{ \pm} }+ 	\pr{+,\widetilde{ \pm} }$, and for each we define an innovations akin to \cref{Eq::HomInnovations}.  
Following the same procedure in \cref{Sec::FockHomodyne}, we obtain \cref{Eq::FockHetSME}.

	%=====================================================
	\subsection{Conditional expectation values} \label{Sec::CondField}
	%=====================================================

The conditional expectation value of a joint operator $\mathcal{O} \in \mathcal{H}_{\rm sys} \otimes \mathcal{H}_{\rm field}$ is
	\begin{align} \label{Eq::ConditionalExpectation}
		\mathbbm{E}[\mathcal{O}(t)|\mRec] \equiv \frac{1}{\pr{\mRec}}\Tr \big[ \condOp \rho_0 \otimes \op{N_\xi}{N_\xi} \condOp\dg \mathcal{O} \big].
	\end{align}
For system operators, $\mathcal{O}= X(t_0) = X \otimes I_{\rm field}$, performing the field trace allows the conditional expectation value to be written in terms of the reduced system state that arises from the solutions of the SMEs above,  
	\begin{align} \label{Eq::ConditionalExpectationX}
		\mathbbm{E}[X(t)|\mRec] = \Tr [ \rho_{\rm sys}(t) X].
	\end{align}

Conditional expectation values for general field operators do not have as simple a reduction as in \cref{Eq::ConditionalExpectationX}.
However, for the output quantum noise increments given by \cref{Eq::IO}, $\mathcal{O}= I_{\rm sys} \otimes dB_t^{\rm out}$ and $\mathcal{O}= I_{\rm sys} \otimes d\Lambda_t^{\rm out}$, the conditional expectation values are readily calculated for Fock-state input. 
Using the solutions to the Fock-state SMEs, conditional expectation values of field observables can be calculated. 
Inserting the Hermitian output field observables, $d\Lambda_t^{\rm out}$ and $dQ^{\rm out}_\phi$, into \cref{Eq::ConditionalExpectation}, 
	\begin{subequations} \label{Eq::ProbFieldRelations}
	\begin{align}
		\mathbbm{E}[d\Lambda_t^{\rm out}|\mRec] &= \pr{J}  \label{Eq::dLambdaCond}, \\
		\mathbbm{E}[dQ_\phi^{\rm out}|\mRec] &= dt \HomCurr  \label{Eq::dQCond} .
	\end{align}
	\end{subequations}
These quantities have indeed already appeared in the probabilities for the measurement outcomes, \cref{Eq::PhotonJumpProb} and \cref{Eq::HomProb}.
The conditional statistics for infinitesimal measurements are fully determined by four operators: the reduced system state $\rho_{N,N}$ along with the three auxiliary operators, $\rho_{N-1,N}$, $\rho_{N,N-1}$, and $\rho_{N-1,N-1}$. 
This is a straightforward consequence of the fact that such measurements are described in a basis with at most one photon. 

As an example, we may be interested in the conditional photon counting statistics given homodyne measurements up to time $t$.  
We first solve \cref{Eq::FockHomSME} for a given homodyne record $\mRec$, then use the solutions to calculate $\pr{J}$, which is equivalent to  $\mathbbm{E}[d\Lambda_t^{\rm out}|\mRec]$ according to \cref{Eq::dLambdaCond}.

	%=====================================================
	\subsection{System-field entanglement}
	%=====================================================

As the quantum system and input Fock state interact they become entangled, and at intermediate times signatures of this entanglement are present in the reduced system state. 
Although the measurements disentangle the portion of the field that has been detected, entanglement between the quantum system and future field persists. 

Indeed, even when the input system state is pure, a trace over the field yields a \emph{mixed} reduced state since the system has become correlated with the unmeasured field, as was studied recently for a two-level atom interacting with a single-photon Fock state \cite{DaeiShei13}.  This is in contrast to input field states that factorize temporally, where the reduced conditional state remains pure if the field is measured with perfect efficiency. Thus, a stochastic Schr\"{o}dinger equation for a pure-state wavefunction does not apply for Fock-state input, and one is required to use SMEs.

%=====================================================
\section{Generalizations} \label{Sec::Generalizations}
%=====================================================		
	
	%=====================================================
	\subsection{Superpositions and mixtures of Fock states in the same wave packet} \label{Sec::ComboFockStates}
	%=====================================================

In this section, we generalize the Fock-state SMEs to input states in superpositions and mixtures of Fock states.  Since the Fock states, $\ket{n_\xi}$, form a complete basis in the temporal mode $\xit$, they can be used to construct any state in that wave packet.  For an initial field state 
	\begin{align}  \label{Eq::CombinationFieldInput}
		\rho_{\rm field}(t_0) = \sum_{m,n} c_{m,n} \op{m_\xi}{n_\xi},
	\end{align}
there are two modifications to the Fock-state SMEs presented in Sec. \ref{Sec::SMEFock}. First, the reduced system state is constructed from the $\rho_{m,n}(t)$ using the coefficients $c_{m,n}$. Second, this reduced system state changes the form of the measurement probabilities. However, the coupled SMEs are \emph{identical} to those for pure Fock-state input. Below we discuss the details.

Given an input field with the form of \cref{Eq::CombinationFieldInput}, the reduced system state at time $t$ is
	\begin{align} \label{Eq::RhoTotal}
		\rho_{\rm sys}(t) = \sum_{m,n} c_{m,n} \rho_{m,n}(t),
	\end{align}
where $\rho_{m,n}(t)$ are solutions to the respective Fock-state SMEs in Sec. \ref{Sec::SMEFock} (photon counting, homodyne, or heterodyne). That is, the form of the SMEs that couple the $\rho_{m,n}$ operators is not modified for superpositions and mixtures of input Fock states. However, the diagonal elements and off-diagonal coherences in \cref{Eq::CombinationFieldInput} do affect measurements of the output fields by modifying the conditional probabilities, $\pr{R_t}$, that normalize the post-measurement matrices $\rho_{m,n}(t)$.

For the Fock-state photon-counting SME the probability of detecting a photon at time $t$ becomes
	\begin{align} \label{Eq::JumpProbSuperposition}
		\pr{J}  = & dt \sum_{m,n}  c_{m,n} \Tr \big[ L\dg L \rho_{m,n} +  \sqrt{m} \xit L\dg S \rho_{m-1,n} \nn \\
		& + \sqrt{n} \xits S\dg L \rho_{m,n-1} + \sqrt{mn} |\xit|^2  \rho_{m-1,n-1} \big], 
	\end{align}
with the probability of vacuum detection given by $\pr{\emptyset} = 1 - \pr{J}$.  The photon-counting SME for superpositions/mixtures of Fock states is found by using the detection probability, \cref{Eq::JumpProbSuperposition}, in either form of the Fock-state photon-counting SME, \cref{Eq::FockPCSME} and \cref{Eq::FockPCSMEInnovations}. The reduced system state is found by combining the solutions according to \cref{Eq::RhoTotal}.

For homodyne detection the probabilities $\pr{\pm}$ and conditioning map $\Hop{\phi}$, \cref{Eq::HomProb}  and \cref{Eq::Hop}, involve the modified expected quadrature current,
	\begin{align} \label{Eq::QuadCurrentSuperposition}
		\HomCurr = & \sum_{m,n} c_{mn} \Tr \big[ (e^{-i \phi}L + e^{i \phi} L\dg ) \rho_{m,n}  \\
		&+ e^{-i \phi}\sqrt{m} \xit S \rho_{m-1,n} + e^{i \phi}\sqrt{n} \xits S\dg \rho_{m,n-1}  \big]. \nn 
	\end{align}
The homodyne SME for superposition/mixtures of Fock states is found by using the modified $\HomCurr$, \cref{Eq::QuadCurrentSuperposition}, in the Fock-state homodyne SME, \cref{Eq::FockHomSME}, and then combining the solutions according to \cref{Eq::RhoTotal} to get the reduced system state. The Fock-state heterodyne SME, \cref{Eq::FockHetSME}, is modified similarly. 

In Appendix \ref{Appendix::HeiSMEs} we give the Heisenberg-picture form of the Fock-state SMEs for the general input fields in \cref{Eq::CombinationFieldInput}.

%=====================================================	
	%%% FIGURE 2: Single trajectories %%%
	\begin{figure*}[t]
	\includegraphics[width=\hsize]{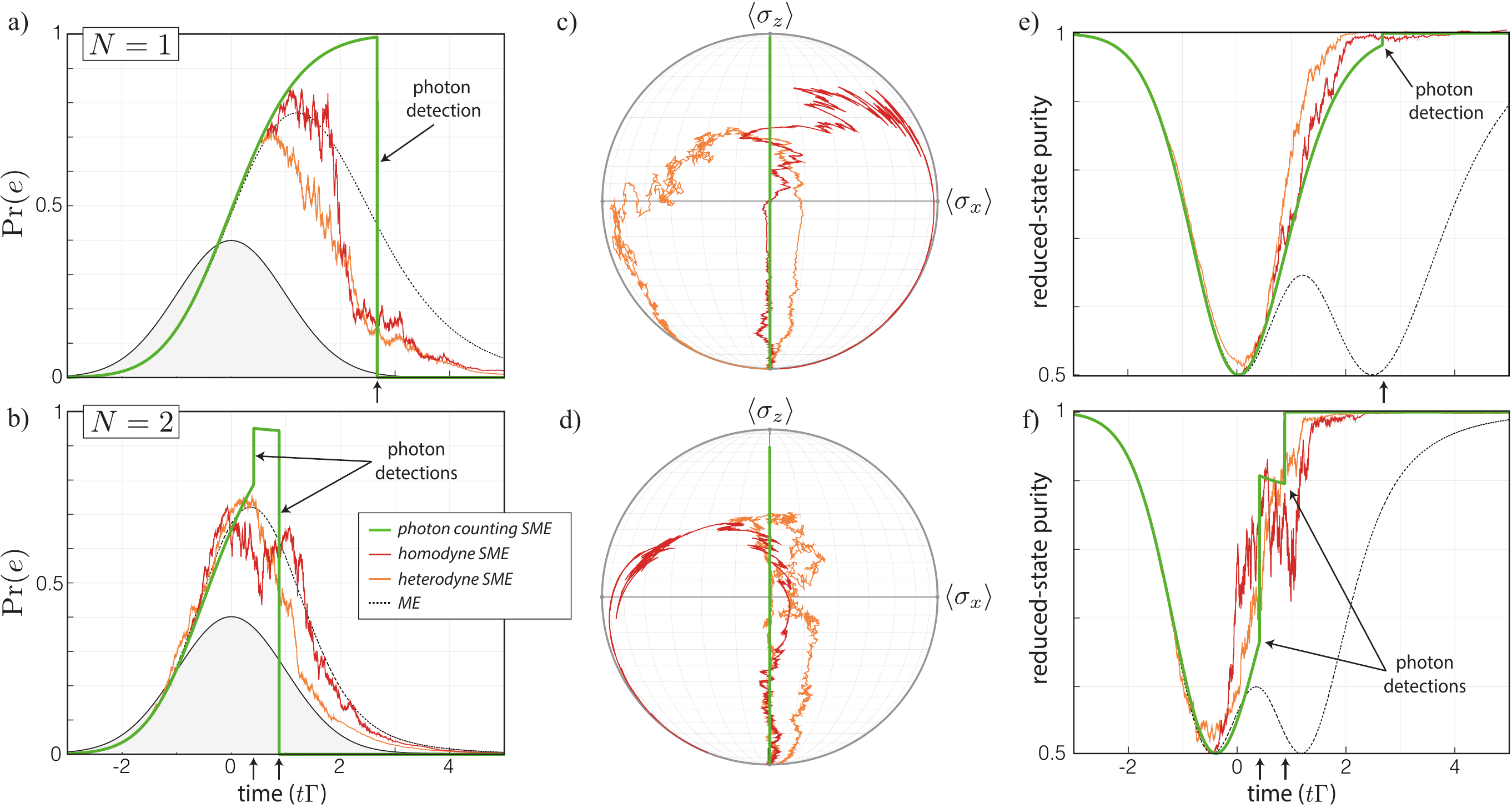}
	 \caption{Single trajectories for a two-level atom interacting with propagating $N=1$ (top row) and $N=2$ (bottom row) Fock states when the output fields are subject to continuous photon counting, homodyne, and heterodyne measurements. Detection times are indicated for the photon-counting trajectory. Unconditional dynamics, calculated from the Fock-state master equation, are shown. (a--b) Conditional excitation probability $\pr{e} =\bra{e} \rho_{\rm sys}(t) \ket{e}$. The input Gaussian wave packet $|\xi(t)|^2$ is shown (thin black filled grey), where $\xi(t)$ is given by \cref{Eq::GaussianWavepacket} with $\Delta_\omega/\Gamma = 1$ and $t_0 = 0$. (c--d) Projection of the Bloch sphere onto the $xz$-plane showing the trajectories of the conditional Bloch vector. For heterodyne detection measurement of both quadratures drives the Bloch vector out of the $xz$-plane (not shown).(e--f) Conditional reduced-state purity, $\Tr [\rho^2_{\rm sys}(t)]$, for the same trajectories.} \label{Fig::SingleTraj}
	\end{figure*}
%=====================================================

	%=====================================================
	\subsection{Imperfect detection} \label{Sec::Imperfect detection}
	%=====================================================

To model detectors of imperfect quantum efficiency $\eta$ ($0 \leq \eta \leq 1$) the Fock-state SMEs and unconditional Fock-state master equations are combined with respective weights $\eta$ and $1-\eta$ \cite{WiseMilb10}.  For photon counting, the standard-form Fock-state photon-counting SME becomes
	\begin{align}
		d\rho_{m,n}(t) = &dN_\eta d\rho_{m,n}(t|J) + \eta d\rho_{m,n}(t|\emptyset) \\
			&+ dt(1-\eta) \mathcal{D}_{L}[\rho_{m,n}]. \nn
	\end{align}
The probability $\pr{J}$ is a statement about the \emph{output field}---it is the probability that a photon arrives at the detector. However, the fact that an imperfect detector may not register the photon is captured by the modified conditional expectation value, $\mathbbm{E}[dN_\eta|\mRec] = \eta \pr{J}$.  Inserting this relation in the photon-counting innovations, \cref{Eq::PCInnovations}, the transformation to the innovations-form SME is straightforward. For homodyne and heterodyne detection, the Fock-state SMEs are obtained by modifying the innovations in Eqs. (\ref{Eq::FockHomSME}) and (\ref{Eq::FockHetSME}) according to $d\mathcal{J}_\phi(t) \rightarrow \sqrt{\eta}d\mathcal{J}_\phi(t)$ \cite{WiseMilb10}.

	%=====================================================
	\subsection{Additional decoherence}
	%=====================================================

Additional decoherence channels can be included by amending the Fock-state SMEs. For each $\rho_{m,n}$ the actions of the Lindblad superoperators corresponding to the decoherence channels are added to the differential maps $d\rho_{m,n}(t)$.
When the dissipation arises from heat baths, it can be derived explicitly by including additional modes in the evolution unitary, \cref{Eq::InfinitesimalUnitary}, and then tracing them out. For example, a stationary thermal bath with mean photon number $\expt{n}$ that couples linearly to the system via the operators $\tilde{L}$ and $\tilde{L}\dg$ introduces additional Lindblad terms the equations of motion for $\rho_{m,n}(t)$. Specifically, to each $d\rho_{m,n}(t)$ the following terms are added, 
	\begin{align}
		%\sqrt{\expt{n} + 1} \mathcal{D}_{\tilde{L}}[\rho_{m,n}] + \sqrt{\expt{n}} \mathcal{D}_{\tilde{L}\dg}[ \rho_{m,n}],
		(\expt{n} + 1) \mathcal{D}_{\tilde{L}}[\rho_{m,n}] + \expt{n} \mathcal{D}_{\tilde{L}\dg}[ \rho_{m,n}],
	\end{align}
where the first describes decay and the second incoherent thermal driving.

	%=====================================================
	\section{Example: conditional dynamics of a two-level atom} \label{Sec::TLSexample}
	%=====================================================

%%=====================================================	
%%	%%% FIGURE 3: Ensemble-averaged quantitites %%%
	\begin{figure*}[t]
	\includegraphics[width=\hsize]{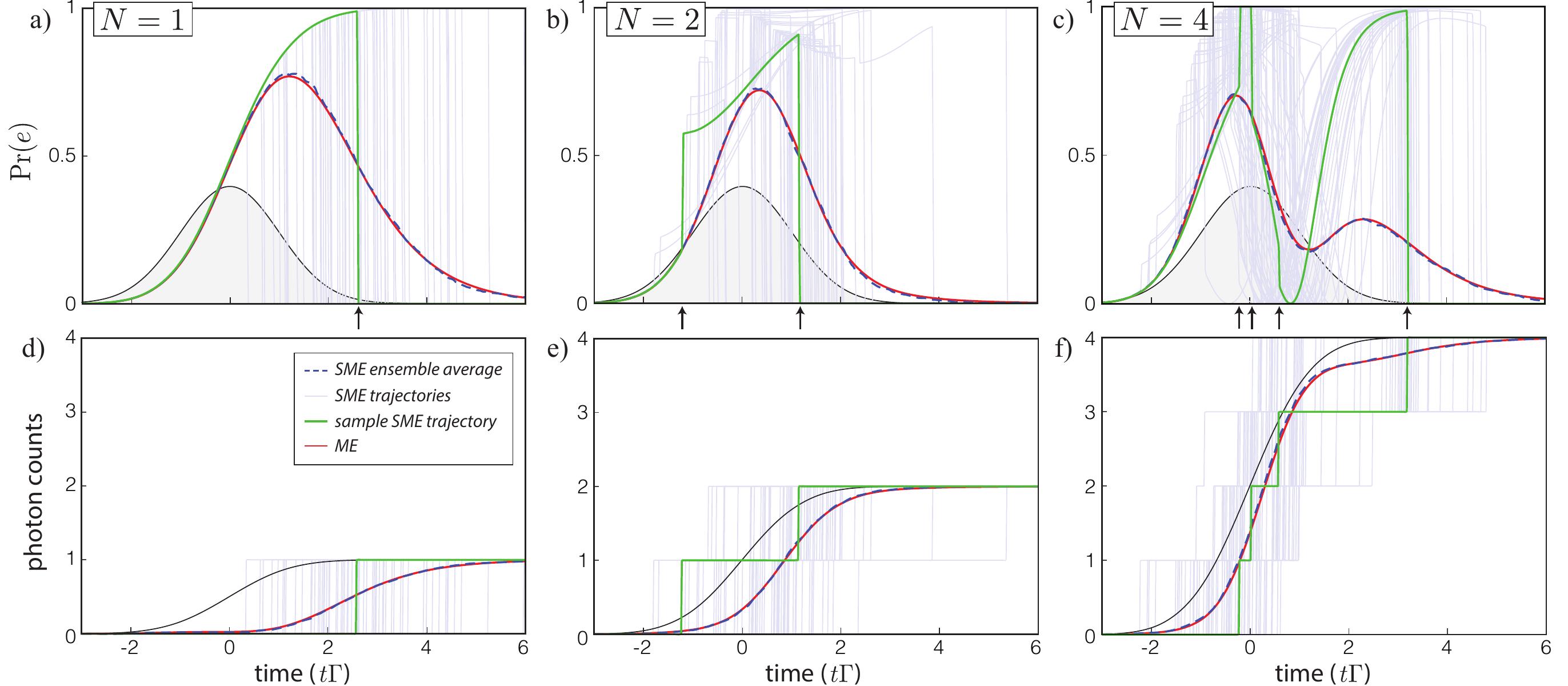}
	 \caption{Ensemble-averaged expectation values for a two-level interacting with Fock states with $N = 1$ (left column), $N = 2$ (middle column) $N = 4$ (right column) photons for continuous photon counting measurements. 1000 trajectories were simulated, 60 of which are shown. For each $N$ a single example trajectory is highlighted (solid curve exhibiting jumps) with times of detection indicated by black arrows on the time axis. The ensemble average over trajectories converges to the unconditional Fock-state master equation. (a--c) Excitation probability $\pr{e}$. The input Gaussian wave packet $|\xi(t)|^2$ is shown (thin black filled grey), where $\xi(t)$ is given by \cref{Eq::GaussianWavepacket} with $\Delta_\omega/\Gamma = 1$ and $t_0 = 0$.  (d--f) Total photon counts as a function of time. Every trajectory has exactly $N$ photon counts, and the asymptotic ensemble averages and unconditional expectation values approach $N$. Average total photon counts for the input Fock state, $\int_{0}^t dt N |\xi(t)|^2$, is shown in black.} \label{Fig::EnsAvg}
	\end{figure*}
%%=====================================================

As an introduction and guide to using the Fock-state SMEs, we present a brief study of conditional dynamics for a two-level atom with eigenstates $\ket{g}$ and $\ket{e}$. 
The total Hamiltonian ($\hbar = 1)$ for the atom-field system is
	\begin{align} \label{Eq::TotalJCHam}
		H &= H_{\rm atom} + H_{\rm field} + H_{\rm int} .
	\end{align}
The bare Hamiltonian for the atom with transition frequency $\omega_0$ is $H_{\rm atom} =  \smallfrac{\omega_{0}}{2}\sigma_z$, where $\sigma_z = \op{e}{e} - \op{g}{g}$.
The positive-frequency, continuous-mode field is described by a bare Hamiltonian $H_{\rm field} = \int_0^\infty d \omega \, \omega  b\dg(\omega) b(\omega)$, with field operators satisfying $[b(\omega), b\dg(\omega')] = \delta(\omega-\omega')$ . Finally, the atom-field interaction is described by the Hamiltonian \cite{GardZoll04},  
	\begin{align} \label{Eq::IntAtomFieldHam}
		H_{\rm int} = -i \int_0^\infty d\omega \, \kappa(\omega) \big[ \sigma_+  b(\omega) - \sigma_- b\dg(\omega) \big],
	\end{align}
where $\kappa(\omega)$ is the dipole-field interaction strength at frequency $\omega$, and $\sigma_+ = \op{e}{g}$ and $\sigma_- = \op{g}{e}$ are the atomic raising and lowering operators.
Making the usual Markov approximation \cite{GardZoll04, Carm93a}, the coupling strength is linearized around the atomic resonance frequency, $\Gamma \equiv 2 \pi | \kappa(\omega_0)|^2$ and the lower limit of integration in \cref{Eq::IntAtomFieldHam} is extended to $-\infty$.  In an interaction picture with respect to $H_0 =  \frac{ \omega_c}{2} \sigma_z + H_{\rm field}$, where $\omega_c$ is the carrier frequency of the input wave packet, the resulting $(S,L,H)$-operators \cite{CombKercSaro16} that appear in the time evolution operator, \cref{Eq::InfinitesimalUnitary}, are
	\begin{subequations}
	\begin{align}
		S =& \Id_{\rm sys},\\		
		L = & \sqrt{\Gamma} \sigma_-,\\
		H_{\rm sys} = & -\Delta_0 \sigma_z,   
	\end{align}	
	\end{subequations}
with $\sigma_z = \op{e}{e} - \op{g}{g}$ and detuning $\Delta_0 \equiv \omega_c - \omega_0$.  
The atom is probed by a Fock state $\ket{N_\xi}$ [\cref{Eq::Fock}] with resonant carrier frequency ($\omega_c = \omega_0$) in a Gaussian wave packet given by
	\begin{align} \label{Eq::GaussianWavepacket}
		\xit = \bigg[  \frac{ (\Delta_\omega/\Gamma)^2}{2\pi}  \bigg]^{1/4} \exp \bigg[ - \frac{(\Delta_\omega/\Gamma)^2}{4} (t - t_0)^2 \bigg].
	\end{align}
The dimensionless spectral bandwidth, $\Delta_\omega/\Gamma$, is defined such that the variance of $|\xit|^2$ is $(\Delta_\omega/\Gamma)^{-2}$.

	%=====================================================
	\subsection{Few-photon Fock states}
	%=====================================================
	
We solve the Fock-state SMEs in Section \ref{Sec::SMEFock} for few-photon input for each of the three measurements: photon counting, homodyne, and heterodyne.  For each measurement type we calculate a single trajectory for the atom initialized in its ground state, $\rho_0 = \op{g}{g}$, and the input field in an $N$-photon Fock state. We numerically integrate the set of coupled equations for $\rho_{m,n}$ and then extract the conditional reduced system state: $\rho_{\rm sys}=\rho_{N,N}$. From $\rho_{\rm sys}$, conditional expectation values are then calculated as usual, see \cref{Eq::ConditionalExpectationX}.

	%=====================================================
	\subsubsection{Single quantum trajectories}
	%=====================================================
	
We begin by examining the conditional excitation probability, \cref{Eq::ConditionalExpectationX}, for $\pr{e} \equiv \tr{\op{e}{e} \rho_{\rm sys}(t) }$ for each of the measurement types. In \cref{Fig::SingleTraj} we plot single trajectories for $N=1$ and $N=2$ input photons. The trajectories for photon counting are of particular interest, as Fock states are eigenstates of total photon number. These trajectories display discrete jumps at the photon detection times---in (a) there is one jump, and in (b) there are two, corresponding to the number of input photons $N$. The vacuum-detection evolution up to the first detection time is deterministic, given by the first line of \cref{Eq::FockPCSME}. At the first detection time the single-photon trajectory returns to zero, indicating a quantum jump to the atom's ground state.  

However, something curious happens for the $N=2$ trajectory in \cref{Fig::SingleTraj}(b): $\pr{e}$ abruptly increases at the first photon detection. Similar quantum jumps up have recently been studied by \citet{BlocMolm17} for a decaying atom. 
 Indeed, the conditional excitation probability may jump up or down depending on time of detection (see  \cref{Fig::SingleTraj}(c)). Further simulations (not shown) indicate that this is a generic feature of photon counting %, present even for input coherent states, and 
arising from interference between input and reradiated fields.  

Insight into the conditional dynamics can be seen in the trajectories of the Bloch vector, $\vec{\sigma} = ( \expt{\sigma_x}, \expt{\sigma_y}, \expt{\sigma_z} )$, shown in \cref{Fig::SingleTraj}(c-d). The Bloch-vector components are  $\expt{\sigma_i(t)} \equiv \Tr [\rho_{\rm sys}(t) \sigma_i]$ for Pauli operators on the pseudospin, 
$\sigma_x = \op{e}{g} + \op{g}{e},  \sigma_y = -i(\op{e}{g} - \op{g}{e})$. 
As pure Fock states have no associated phase, the excitation dynamics for photon counting trajectories lie entirely on the $z$-axis of the Bloch sphere. At the center of the Bloch sphere, where $\pr{e} = 0.5$, the atom is maximally entangled with the field and the reduced state reaches its minimum purity.  For the diffusive cases of homodyne and heterodyne detection, the phases associated with the measurements drive the reduced atomic state off the $z$-axis, as seen in the Bloch-sphere representations of the state in Fig. \ref{Fig::SingleTraj}(a-b). This results in reduced states of higher purity in general. 
Since the atom is prepared in a pure state, atom-field entanglement is revealed by the reduced-state purity of the atom, $\Tr [\rho^2_{\rm sys}(t) ]$,  plotted in \cref{Fig::SingleTraj}(e-f). 

%=====================================================	
%%% FIGURE 4: Coherent State Approximation %%%
\begin{figure*}[t]
	\includegraphics[width=\hsize]{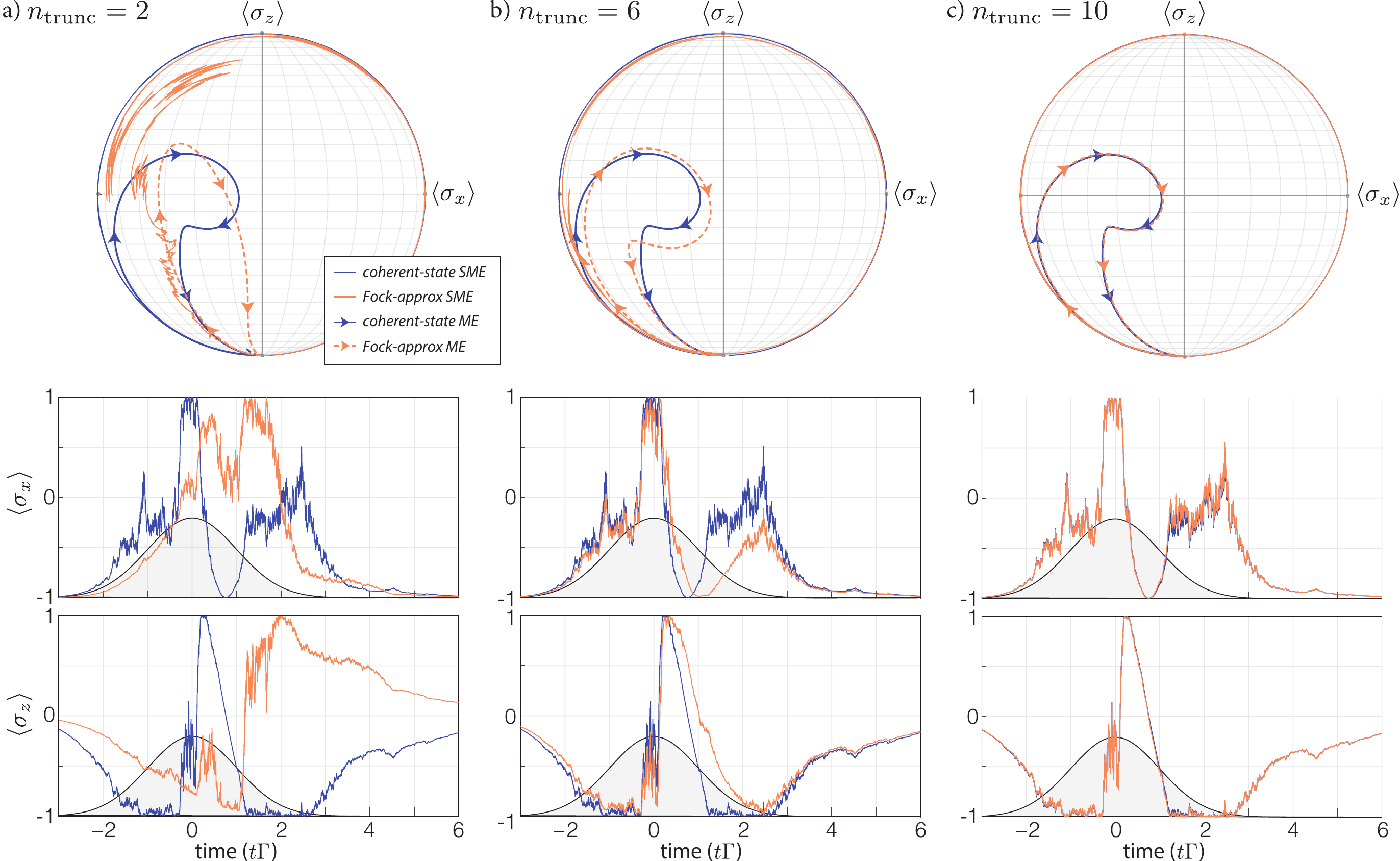}
	 \caption{Trajectories for a two-level atom interacting with finite approximations (red) to a coherent-state wave packet (blue) of increasing precision (Fock-state truncation level $n_{\rm trunc})$, when the output fields are subject to homodyne detection. The coherent state is prepared with $\expt{n} = 5$ photons. Convergence of a single trajectory is shown under the same noise realization (simulated measurement record).  (a-c) Projection of the Bloch sphere onto the $xz$-plane showing the trajectories of the conditional Bloch vectors with respective Fock-state approximations. The approximations are truncated at $n_{\rm trunc} = \{ 2, 6, 10 \}$, which account for $\{0.04, 0.62, 0.97\}$ of the total photons in the coherent state. For comparison, the Bloch vectors for the coherent-state and Fock-state-approximation master equations are shown. Below each Bloch sphere are the dynamics of $\expt{\sigma_x(t)}$ and $\expt{\sigma_z(t)}$. For reference the input Gaussian wave packet $|\xi(t)|^2$ is shown in black filled grey, where $\xi(t)$ is given by \cref{Eq::GaussianWavepacket} with $\Delta_\omega/\Gamma = 1$ and $t_0 = 0$.   } \label{Fig::CohStateApprox}
\end{figure*}
%=====================================================	

	%=====================================================
	\subsubsection{Ensemble-averaged quantities}
	%=====================================================

In early work by \citet{DaliCastMolm92}, quantum trajectories for vacuum input fields were used not as a description of a single-shot continuous measurement, but rather as a tool to efficiently simulate ensemble-averaged, unconditional evolution. This relies on the fact that as the number of trajectories becomes large, the ensemble average over quantum trajectories approaches the unconditional master equation (ME). Convergence to the ME is similarly true for Fock-state input fields, as demonstrated in \cref{Fig::EnsAvg}. For each input field preparation with $N= \{ 1,2,4 \}$ photons, we simulated 1000 trajectories of the photon-counting SMEs. In the top row, we illustrate agreement of the ensemble-averaged SME and the ME for $\pr{e}$.

Below, in the second row of \cref{Fig::EnsAvg}, we plot total photon counts as a function of time. In any particular trajectory the photon detection times are random, but as $t \rightarrow \infty$ the total number of detections is equal to the number of input photons. The ensemble average over trajectories approaches the unconditional integrated photon flux, $\int_0^t dt'   \expt{d\Lambda^{\rm out}_{t'}}$. This is directly calculated from the input-output field relation, \cref{Eq::dLamOut}, using the Fock-state MEs as described in Ref. \cite{BaraCookBrai12}. 

%=====================================================
\subsection{Fock-state approximation to a coherent state}
%=====================================================

A continuous-mode coherent state with amplitude $\alpha(t) = \alpha_0 \xit$, with peak amplitude $\alpha_0$ \footnote{The overall phase could be included in the wave packet $\xit$, but we choose here to follow convention.}, can be expanded in the basis of Fock states as \cite{Loud00} 
	\begin{align} \label{Eq::CohStateFockBasis}
		\ket{\alphatot} = e^{- |\alpha_0|^2/2} \sum_{n=0}^\infty \frac{ (\alpha_0)^n}{ \sqrt{n!} } \ket{n_\xi}.
	\end{align}
The total mean photon number is $\expt{n} = |\alpha_0|^2$ and peak input photon flux is $\expt{n} \,  \mbox{max}  |\xit|^2 $. A finite approximation to $\ket{\alphatot}$ is found by truncating the Fock expansion at chosen photon number $n_{\rm trunc}$ and renormalizing.  

We return to the two-level atom in Sec. \ref{Sec::TLSexample}.  
Here, we consider the case where the atom is probed by a coherent-state wave packet $\ket{\alphatot}$ with amplitude $\alpha_0 = \sqrt{5}$ corresponding to $\expt{n} = 5$, and the output fields are measured via continuous homodyne detection.  The wave packet is Gaussian, \cref{Eq::GaussianWavepacket}, with $\Delta_\omega/\Gamma = 1$. 
In \cref{Fig::CohStateApprox} we compare the conditional coherent-state dynamics to those for Fock-state approximations of increasing truncation $n_{\rm trunc}$ in \cref{Eq::CohStateFockBasis}. 
The input state is a superposition of Fock states, so the reduced system state is given by \cref{Eq::RhoTotal}.
Under identical measurement records, the approximate Fock-state conditional expectation values converge to the coherent-state values as the truncation level increases, as seen in columns (a--c).  
Shown for comparison are the unconditional ME dynamics for the Fock-state approximation \cite{BaraCookBrai12, Bara14}, which likewise converge.  

For coherent-state input, the conditional state of the atom remains pure at all times and its Bloch vector is confined to the surface of the Bloch sphere. Entanglement generated between the atom and field in each infinitesimal time interval is immediately recovered by the measurement, which projects the joint state into a tensor-product state. This is evident from the coherent-state trajectory, which traces out the boundary of the Bloch sphere in the $xz$-plane. When the truncation level of the the Fock-state approximations is low, Fig \ref{Fig::CohStateApprox}(a) for example, the conditional Bloch vector enters the interior and the reduced-state becomes mixed, signaling residual entanglement with the future, unmeasured field.

%=====================================================
\section{Conclusion}  \label{Sec::Conclusion}
%=====================================================

We have presented the stochastic master equations that describe the conditional reduced-state dynamics for a quantum system interacting with a propagating $N$-photon Fock state. Our derivation of the Fock-state SMEs for different detection schemes uses Kraus operators and a temporal decomposition of the input fields, rather than the methods of quantum filtering theory~\cite{BoutHandJame07}. 

A method complementary to our coupled SMEs exists, which treats the input Fock state as arising from a cascaded source. This was originally suggested for a single photon by \citet{GherElliPell98} and \citet{GougJameNurd11,GougJameNurd12}, and generalized to $N$ photons by \citet{GougZhan15}. 
Further, many alternative approaches have been developed to study quantum interactions with propagating photons. These include path-integral-based master equations \cite{ShiChanCira15}, generalized input-output theory \cite{XuFan15}, direct calculations of the scattering matrix \cite{LeeNohSche15}, time-domain treatment of the joint-state wavefunction \cite{Kosh08, Hong:2013aa, KonyGeaB16}, and diagrammatic methods \cite{RoulScar16}. 
It would be interesting to see if such methods could be extended to describe field measurements and the resulting quantum trajectories.

The Fock-state SMEs presented here can be extended to more general fields. For input states with large mean fields, a Mollow transformation can be used to reduce the number of Fock states required for a faithful simulation ~\cite[see Sec. VII B 1]{CombKercSaro16}. By choosing a temporal-mode basis, as was done for Fock-state MEs in Refs. \cite{BaraCookBrai12, Bara14}, our methods can be straightforwardly extended to more general $N$-photon states whose spectral density functions do not factorize \cite{RohdMaueSilb07}.  Recently, using a non-Markovian embedding in the Heisenberg picture, SMEs for such states were derived by Song \emph{et al.} in Ref. \cite{SongZhanXi16}. 

There are a number of applications of our theory. Single-photon SMEs have already been used to study conditional phase shifts on a cavity \cite{CarvHushJame12}, nondestructive photon detection \cite{SathTornKock14}, and conditional excitation probabilities ~\cite{GougJameNurd12,Dabr16}. Two-mode, two-photon quantum trajectories have been used to study an effective beamsplitter interaction using a mechanical resonator \cite{BasiMyerComb16} and excitation probabilities~\cite{DongZhanAmin16}. It is forseeable that our formalism could enable the study of high-speed quantum feedback control~\cite{WiseMilb10, Liu:2016aa} with nonclassical fields. Further, the Fock-state SMEs could allow for novel state preparation using time-resolved post-selection, where the number of photons detected as well as the \emph{times of detection} are design parameters. Finally, the techniques of \cref{Sec::ComboFockStates} allow for the direct simulation of SMEs for exotic field states by expressing them in a Fock basis.  

Another broad application of the Fock-state SMEs is the study of quantum networks with nonclassical inputs. Carmichael and Gardiner~\cite{Carm93,Gard93} developed the theory of \emph{cascaded quantum systems} for situations where one quantum system is driven by the output of another. Recently, this theory was expanded to include the scattering operator $S$ and formalized into the \emph{SLH framework} for quantum networks~\cite{GougJame09}. For a set of $n$ cascaded quantum systems, the theory describes rules to compose the separate Hamiltonians $H^{(n)}$, linear coupling operators $L^{(n)}$, and scattering operators $S^{(n)}$ for  
into a set of new $(S,L,H)_{\rm cas}$ operators that describes the joint system as a whole. An introduction can be found in Ref.~\cite{CombKercSaro16}.  The cascaded coupling operators, $(S,L,H)_{\rm cas}$, are used directly in the Fock-state SMEs above. Using the results in \cref{Sec::ComboFockStates}, our formalism allows simulating an arbitrary field state input to the network. 

{\em Acknowledgements:} The authors would like to thank (in alphabetical order):  Sahar Basiri-Esfahani, Agata Bra\'nczyk,
 Carl Caves%("sometimes you just have to cut the cord")
, Robert Cook, Ivan Deutsch, Bixuan Fan, Julio Gea-Banacloche, John Gough, Michael Jack, 
Matthew James, Zhang Jiang, G\"oran Johansson, Anton Kockum, Casey Myers, Gerard Milburn, 
Hendra Nurdin, Nicolas Quesada, Sankar Sathyamoorthy, Tom Stace, Lars Tornberg, and Howard 
Wiseman for discussions on related topics over the last several years.  BQB thanks the Perimeter 
Institute for financial support during several research visits. This research was supported in part by 
the ARC Centre of Excellence in Engineered Quantum Systems (EQuS),
Project No. CE110001013. 
BQB and JC acknowledge financial support from NSF Grant No. 
PHY-0969997, No. PHY-0903953, No. PHY-1005540, and No. PHY-1521016, ONR Grant No. 
N00014-11-1-008, and AFOSR Grant No. Y600242. 
JC was supported by the Australian Research Council via DE160100356 and the Perimeter Institute 
for Theoretical Physics. Research at Perimeter Institute is supported
by the Government of Canada through the Department
of Innovation, Science and Economic Development
Canada and by the Province of Ontario through
the Ministry of Research, Innovation and Science.
\\

%=====================================================	
%%% BIBLIOGRAPHY	
%

%=====================================================

%=====================================================		
%%%  APPENDIX  %%%
\begin{appendix}	
%=====================================================

%=====================================================
\section{Temporal decomposition of Fock states} \label{Appendix:TempDecomp}
%=====================================================	

Here we expand on the temporal decomposition of Fock states within the wave packet $\xit$. With respect to a time $t$, the wave packet creation operator decomposes as
	\begin{align}
		B\dg\pis(\xi) &= \int_{t_0}^t ds \xi_s b\dg (s) +  \int_{t}^\infty ds \xi_s b\dg (s) \\
		&= B\dg\pis(\xi) + B\dg_{[t}(\xi) ,
	\end{align}
where by definition $[B\dg\pis(\xi), B\dg_{[t}(\xi)] = 0$.
Inserting this decomposition into \cref{Eq::Fock} and expanding the product, a Fock state can be written
	\begin{align}
		\ket{n_\xi} 
		& = \frac{1}{\sqrt{n!} } \sum_{k=0}^n \binom{n}{k} \left[ B_{t)}^{\dagger } (\xi) \right]^k \left[ B\dg_{[t} (\xi) \right]^{n-k}  \mmvac  \label{Eq::FockStateDecomp1}
	\end{align}
with binomial coefficient
	\begin{align}
		\binom{n}{k} \equiv \frac{n!}{k!(n-k)!}.
	\end{align}
The fact that \cref{Eq::FockStateDecomp1} does not factorize in time indicates temporal correlations between photons in the past and future time intervals. 
Applying a temporally decomposed wave packet creation operator to vaccuum generates \emph{unnormalized} Fock states over a time interval (chosen here to be the past interval $[t_0, t)$, for illustration)
	\begin{align} \label{Eq::FockUnnorm}
		\ket{ \overline{n} \pis } = \frac{1}{ \sqrt{n!} }  \left[ B\pis^{\dagger } (\xi) \right]^n \mmvac .
	\end{align}
They are unnormalized,
	\begin{align} \label{Eq::FutureFieldTrace}
		\ip{ \overline{n}_{t)} }{ \overline{n}_{t)} } &= [ 1-w(t) ]^n \\
		\ip{ \overline{n}_{[t} }{ \overline{n}_{[t} } &= [w(t)]^n .
	\end{align}
with
	\begin{align}
		w(t) \equiv \int_t^\infty dt |\xit|^2 \leq 1,
	\end{align}
because the division of $\xit$ creates two temporal modes that are not individually square-normalized. Then, we can express \cref{Eq::FockStateDecomp1} as
	\begin{align} \label{Eq::FockStateDecomp2}
		\ket{n_\xi} & = \sum_{k=0}^n \sqrt{ \binom{n}{k} } \ket{ \overline{k}_{t)} } \otimes \ket{ \overline{n-k}_{[t} }.
	\end{align}
Using \cref{Eq::FutureFieldTrace} the past and future Fock states can be normalized,
	\begin{align} \label{Eq::FockStateDecomp2}
		\ket{n_{t)}} &=  [1 - w(t)]^{-\frac{n}{2}} \ket{ \overline{n}_{t)} } \\
		\ket{n_{t)}} &=  [w(t)]^{-\frac{n}{2}} \ket{ \overline{n}_{t)} }
	\end{align}	
so that the temporal decomposition becomes
	\begin{align} \label{Eq::FockStateDecomp2}
		\ket{n_\xi} & = \sum_{k=0}^n \sqrt{ \binom{n}{k} } \sqrt{ [1 - w(t)]^{k} [w(t)]^{n-k} } \ket{ k_{t)} } \otimes \ket{ n-k_{[t} }.
	\end{align}

	%=====================================================
	\subsection{Infinitesimal decomposition}
	%=====================================================
	
Following the same procedure, we decompose the Fock state according to the three-interval temporal decomposition in \cref{Eq::WavePacketTempDecomp}. However, we choose to combine the past and future wave packet creation operators, only making the distinction with the current time interval $[t, t+dt)$. To this end we define $\cutout{B}\dg \equiv B\dg\pis(\xi) + B\dg\fis(\xi).$ Then \cref{Eq::NPhotonDecomposition} can be written,	
	\begin{align}
		\ket{n_\xi} & = \frac{1}{\sqrt{n!} } \left[  \xit dB\dg_t  + \cutout{B}\dg \right]^n \mmvac \nn \\
		& = \frac{1}{\sqrt{n!} } \sum_{k = 0}^n \binom{n}{k} \big( \xit dB\dg_t \big)^{k} \big( \cutout{B}\dg \big)^{n-k}  \mmvac \nn \\
		& = \left[\frac{1}{\sqrt{n!}} \big( \cutout{B}\dg \big)^n + \sqrt{n}\xit dB_t\dg \frac{1}{\sqrt{(n-1)!}} \big( \cutout{B}\dg \big)^{n-1} \right] \mmvac \nn \\
		& = \ket{0\dt} \otimes \ket{\cutout{n}} + \sqrt{n \, dt} \xit \ket{1_t} \otimes \ket{\cutout{n-1}}. \label{Eq::KetRelationAp}
	\end{align} 	
The third line follows from the It\={o} relation $dB_t\dg dB\dg_t = 0$, such that the only nonvanishing terms are $k=0$ and $k=1$, and in the last line we have used the definition of the infinitesimal single-photon state,  \cref{Eq::InfSinglePhoton}. Equation (\ref{Eq::KetRelationAp}) is the relative-state decomposition in \cref{Eq::KeyRelation}.
The unnormalized Fock states $\ket{ \cutout{n} } = (1/ \sqrt{n!})   \big[ \cutout{B}^{\dagger } \big]^n \ket{0\pis} \otimes \ket{0\fis} $ are defined on the joint past-future Hilbert space, $\mathcal{H}\pis \otimes \mathcal{H}\fis$, which excludes the current interval. From \cref{Eq::KetRelationAp} the $\ket{ \cutout{n} } $ are clearly equivalent to the definition in \cref{Eq::CutoutProject}. Their inner product, $\ip{\cutout{n}}{\cutout{n}} = 1 - n \, dt |\xit|^2$, can be worked out directly or by iteration as in \cref{Eq:ben_norm}.

%=====================================================
\section{Heisenberg-picture Fock-state SMEs} \label{Appendix::HeiSMEs}
%=====================================================	

In some cases it proves useful to condition \emph{operators}, rather than quantum states, on the continuous measurement record $\mRec$. In the mathematical quantum filtering literature \cite{BoutHandJame07}, this is the preferred picture. 
Given the initial joint state $\rho_0 \otimes \rho_{\rm field}$, the conditional expectation of a system operator $X(t_0) = X \otimes I$ \cite{BoutHandJame07}, is 
	\begin{align} \label{Eq::HeisenbergCondOp}
		\piC{}{X} &\equiv \frac{1}{\pr{\mRec}} \Tr_{\rm field} \Big[ \rho_{\rm field}  \condOp\dg X \condOp  \Big].
	\end{align}
Conditional expectation values are calculated as $\mathbbm{E}[X|\mRec] = \Tr_{\rm sys}\big[ \rho_0 \piC{}{X} \big]$. For input fields of the form of \cref{Eq::CombinationFieldInput}, the conditional operator in \cref{Eq::HeisenbergCondOp} at time $t$ is found by solving a set of coupled SMEs for the Fock-state conditional operators,
	\begin{align} 
		\piC{m,n}{X} &\equiv \frac{1}{\pr{\mRec}} \bra{m_\xi}  \condOp\dg X \condOp \ket{n_\xi},
	\end{align}
and then reconstructing $\piC{}{X}$ from the coefficients:
	\begin{align} 
		\piC{}{X} = \sum_{m,n} c_{m,n}^* \piC{m,n}{X}.
	\end{align}
Since $\piC{}{X}$ is an operator, conditional expectation values are found by taking the trace with the initial system state, $[\piC{m,n}{X}]\dg = \piC{n,m}{X\dg}$.	
The SMEs for $\pi_{m,n}[X] $ \cite{GougJameNurd12,SongZhanXi16} can be extracted directly from the Fock-state SMEs above using the cyclic property of the trace on each term. For example, multiplying a term $A \rho_{m,n}(t) B$ by $X$ and taking the trace over system and field one can show,
	\begin{align}
		\Tr & [ A \rho_{m,n}(t) BX] = \Tr_{\rm sys} \big[ \rho_0 \piC{n,m}{ B X A} \big]. 
	\end{align}
Repeating this procedure on each of the Fock-state SMEs in \cref{Sec::SMEFock} and identifying the terms in the system trace yields the SMEs in Heisenberg-form.

\begin{widetext}
For continuous photon counting the Heisenberg-picture Fock-state SME is
	\begin{align} \label{Eq::HeiFockPC} 
	&d\piC{m,n}{X}  =  \\
	&dt \Big\{ i  \piC{m,n}{ [H_{\rm sys},X] } - \smallfrac{1}{2} \piC{m,n}{ \{ L\dg L,X \}_+  } - \sqrt{m}\xits \piC{m-1,n}{ S\dg L X } - \sqrt{n}\xit \piC{m,n-1}{ X L\dg S } \Big\} + \pr{J} \piC{m,n}{X} \nn \\
		&  + d N \left (  \frac{ \piC{m,n}{ L\dg X L } + \sqrt{m}\xits \piC{m-1,n} {S\dg X L} +  \sqrt{n}\xit \piC{m,n-1}{ L\dg X S } + \sqrt{mn }|\xit|^2 \piC{m-1,n-1}{ S\dg XS} }{\pr{J}/dt} -\pi_{m,n}(X)\! \right ) \nn 
	\end{align}
with probability of photon detection
	\begin{align}
		\pr{J} = & dt \sum_{m,n}  c^*_{m,n} \Tr_{\rm sys} \Big[ \rho_0 \big\{ \piC{m,n}{L\dg L} +  \sqrt{n} \xit \piC{m,n-1}{L\dg S} + \sqrt{m} \xits \piC{m-1,n}{S\dg L} 
		  + \sqrt{mn} |\xit|^2  \piC{m-1,n-1}{I}  \big\} \Big]. 
	\end{align}	
For homodyne detection the Heisenberg-picture Fock-state SME is,
	\begin{align} \label{Eq::HeiFockHom}
		d & \piC{m,n}{X}  = dt \Big\{ i\piC{m,n}{[H_{\rm sys},X]} +\piC{m,n}{ \mathcal{D}\dg_{L}[X]} \\
		 & + \sqrt{m}\xits\piC{m-1,n}{ S\dg  [ X, L]  } +  \sqrt{n}\xit \piC{m,n-1}{  [ L\dg,X]S  } \nn 
 		 +\sqrt{mn }|\xit|^2 \piC{m-1,n-1}{S\dg XS - X} \Big\} \nn \\
		 &+ d\mathcal{J}_\phi(t) \Big\{ e^{i \phi} \big( \piC{m,n}{LX}+ \sqrt{m}\xits \piC{m-1,n}{S\dg X} \big)  + e^{-i \phi} \big( \piC{m,n}{X L\dg} + \sqrt{n}\xit \piC{m,n-1}{X S} \big)  - \HomCurr \piC{m,n}{X} \Big\}, \nn 
	\end{align}
where the adjoint-form Lindblad superoperator acting on $X$ is $\mathcal{D}\dg_L[X] \equiv L\dg X L - \half \big( L\dg L X + X L\dg L\big).$ The expected homodyne current is
	\begin{align} 
		\HomCurr = & \sum_{m,n} c^*_{mn} \Tr_{\rm sys} \Big[ \rho_0\big\{ \piC{m,n}{e^{-i \phi}L + e^{i \phi} L\dg }
		+ e^{i \phi}\sqrt{n} \xit \piC{m,n-1}{S} + e^{-i \phi}\sqrt{m} \xits \piC{m-1,n}{S\dg}  \big\} \Big].  
	\end{align}
The Heisenberg-form SME for heterodyne detection follows straightforwardly from the homodyne SME in analogy with \cref{Eq::FockHetSME}

In each case, the SME for conditional operator $\piC{m,n}{X}$ includes terms that couple to other conditional system operators, \emph{e.g.} $\piC{m-1,n}{S\dg X L}$, whose SMEs must be also solved simultaneously. 
\end{widetext}

\end{appendix}

\end{document}